\documentclass{jfm}
\usepackage{lineno,hyperref}
\usepackage{hyperref}
\usepackage{fancyhdr}
\usepackage{amsmath}
\usepackage{mathtools}
\usepackage{amsfonts}
\usepackage{amssymb}
\usepackage{textcomp}
\usepackage{booktabs}                      
\usepackage{multicol}
\usepackage{multirow}
\usepackage{xspace}
\usepackage{nomencl}
\usepackage{subcaption,booktabs}
\usepackage{caption}
\usepackage{float}
\usepackage[dvipsnames]{xcolor}
\usepackage{rotating}
\makenomenclature
\usepackage{tabularx}                      
\usepackage{natbib}
\usepackage{adjustbox}
\usepackage[graphicx]{realboxes}
\usepackage[toc,page]{appendix}
\usepackage{cancel}
\usepackage{graphicx}
\usepackage{subcaption}
\usepackage{epstopdf}
\usepackage{color}

\let\OldS\S
\renewcommand{\S}{\OldS\xspace}

\newcommand{\nsc}[1]{{\color{black}#1}}
\newcommand{\nsl}[1]{{\color{black}#1}}

\newcommand{\revA}[1]{{\color{black}#1}}

\newcommand{\revB}[1]{{\color{black}#1}}

\newcommand{\revC}[1]{{\color{black}#1}}

\shorttitle{Finite-size evaporating droplets in HST}

\shortauthor{Scapin et. al.}

\title{Finite-size evaporating droplets in weakly compressible homogeneous shear turbulence}
\author{Nicolò Scapin\aff{1}
  \corresp{\email{nicolos@mech.kth.se}},
  Federico Dalla Barba\aff{2},
  Giandomenico Lupo\aff{1},
  Marco Edoardo Rosti\aff{3},
  Christophe Duwig\aff{4},
  Luca Brandt\aff{1,5}  
}
\affiliation
{
\aff{1}Department of Engineering Mechanics, Royal Institute of Technology (KTH), Stockholm, Sweden,
\aff{2}Department of Industrial Engineering \& CISAS, University of Padova, Padova, Italy,
\aff{3}Complex Fluids and Flows Unit, Okinawa Institute of Science and Technology Graduate University (OIST), 1919-1 Tancha, Onna-son, Okinawa 904-0495, Japan,
\aff{4}Department of Chemical Engineering, Royal Institute of Technology (KTH), Stockholm, Sweden,
\aff{5}Department of Energy and Process Engineering, Norwegian University of Science and Technology (NTNU), Trondheim, Norway
}

\begin{document}

\maketitle

\begin{abstract} 
We perform interface-resolved simulations of finite-size evaporating droplets in weakly compressible homogeneous shear turbulence. The study is conducted by varying three dimensionless physical parameters: the initial gas temperature over the critical temperature $T_{g,0}/T_c$, the initial droplet diameter over the Kolmogorov scale $d_0/\eta$ and the surface tension, i.e.\ the shear-based Weber number, $We_{\mathcal{S}}$. \\
For the smallest $We_{\mathcal{S}}$, we first discuss the impact on the evaporation rate of the three thermodynamic models employed to evaluate the gas thermophysical properties: a constant property model and two variable-properties approaches where either the gas density or all the gas properties are allowed to vary. Taking this last approach as reference, the model assuming constant gas properties and evaluated with the "1/3" rule is shown to predict the evaporation rate better than the model where the only variable property is the gas density. Moreover, we observe that the well-known Fr{\"o}ssling/Ranz-Marshall correlation underpredicts the Sherwood number at low temperatures, $T_{g,0}/T_c=0.75$. \\
Next, we show that the ratio between the actual evaporation rate in turbulence and that computed in stagnant conditions is always much higher than one for weakly deformable droplets: it decreases with $T_{g,0}/T_c$ without approaching unity at the highest $T_{g,0}/T_c$ considered. This suggests an evaporation enhancement due to turbulence also in conditions typical of combustion applications. \\
Finally, we examine the overall evaporation rate and the local interfacial mass flux at higher $We_{\mathcal{S}}$, showing a positive correlation between evaporation rate and interfacial curvature, especially at the lowest $T_{g,0}/T_c$. 
\end{abstract}


\section{Introduction}\label{sec:intro}
%
%
The understanding of droplets evaporation in turbulent flows is a crucial aspect in many different contexts, such as geophysics and spray combustion to name a few. In the first example, evaporation is central for the formation and evolution of clouds and, more in general, in many of the {\it grand challenges of environmental fluid mechanics} \citep{dauxois2021confronting}. In the second example, droplet evaporation is a precursor of combustion; ensuring that all the liquid vaporizes before chemical reactions occur is fundamental to minimize the pollutants formation and to maximize the efficiency of the entire process \citep{birouk2006current}. More recently, evaporation in turbulence acquired a prominent role also in our understanding of the fluid dynamics aspects of COVID-19 spreading (e.g., droplet generation due to exhalation and airborne dispersion) as described in \cite{mittal2020flow,balachandar2020host,bourouiba2021fluid}. Therefore, several studies have been conducted to understand how relative humidity affects evaporation and condensation \citep{rosti2020fluid,ng2021growth,chong2021extended} as well as on the interaction between droplets and turbulent flows \citep{bourouiba2020turbulent,rosti2021turbulence} with the ultimate goal of improving social distancing guidelines. \\
%
%
Historically, evaporating droplets in turbulence have been the subject of numerous and extensive experimental campaigns which have mainly focused on the flow topology inside the droplets~\citep{wong1992internal,mandal2012internal}, on the \nsc{interaction between droplet dispersion and vapour clouds structure~\citep{de2016dense,villermaux2017fine,sahu2018interaction} and on evaporation enhancement of super-Kolmogorov (i.e., finite-size) droplets~\citep{marti2017evaporation,verwey2018fuel,verwey2020experimental}.} On the computational side, the most common  approach has been  the point particle method \citep{kuerten2016point,maxey2017simulation} for the studies of sub-Kolmogorov droplets in different flow configurations, i.e. forced homogeneous isotropic turbulence \citep{mashayek1998direct,weiss2018evaporating}, homogeneous shear turbulence \citep{mashayek1998droplet,weiss2020evaporating}, turbulent channel flow \citep{russo2014water,bukhvostova2014comparison} and turbulent jets \citep{reveillon2007effects,wang2021direct}. 
The main limitation of this approach is represented by the inherent need of empirical closure equations to model the mass, momentum and energy coupling between the disperse and the continuous phase. \nsc{This feature limits their rigorous applications only to sub-Kolmogorov droplets \citep{elghobashi2019direct} and hinders the possibility to investigate directly aspects such as the reciprocal influence of nearby droplets \citep{lupo2020direct,chong2021extended} or the evaporation enhancement due to turbulence  \cite[which typically occurs for droplets of the order of the Taylor length,][]{birouk2006current}. Moreover, assessing the impact of droplets deformation on the evaporation rate as well as the local interfacial mass flux distribution over the droplet surface is not possible, unless further models are introduced.}
On the other hand, less work has been dedicated to numerical simulations of droplets larger than the Kolmogorov scale and the only few works available 
consider homogeneous isotropic turbulence (HIT). 
Specifically, in \cite{albernaz2017droplet} the authors studied by means of a hybrid Lattice Boltzmann method the deformation and heat transfer of a single droplet with a diameter between $25\eta$ and $40\eta$, with $\eta$ the Kolmogorov length. 
In this set-up, however, little can be said about the mass transfer because  evaporation and condensation compensate for a statistically constant droplet volume. 
Recently, \cite{dodd2021analysis} employed a geometric volume of fluid (VoF) method to study the evaporation of droplets at different volume fractions ($0.01$ $\%$ $\leq\alpha\leq$ $1$ $\%$) with an initial droplet diameter ranging between $4\eta$ and $17\eta$, in order to highlight the limitations of point particle closures for the calculation of the evaporation rate and the semi-empirical correlations of the Sherwood number in absence of mean flow and for \nsc{non-isolated droplets}. \\
The limited number of studies and the need of further understanding of such a complex process motivates us to further investigate finite-size evaporating droplets in homogeneous-shear turbulence (HST). As noted in~\cite{kasbaoui2017algorithm}, this configuration can be regarded as one of intermediate complexity between homogeneous isotropic and non-homogeneous flows and it represents a particularly convenient set-up for two reasons. First, it allows to study shear flows without the complications induced by the presence of the walls. Next, given the intrinsic production of turbulent kinetic energy by the mean shear, turbulence is self-sustained without any external forcing and the flow achieves a statistically stationary condition (SS-HST) \citep{pumir1996turbulence,sekimoto2016direct}. The SS-HST has been recently considered in multiphase flows, e.g. in~\cite{rosti2019droplets} for emulsions and in~\cite{tanaka2017effect,yousefi2020modulation} for turbulence modulation by rigid particles of different shapes. In both cases, the characteristic length of the disperse phase was chosen larger than the Kolmogorov scale. \\
%
%
\nsc{
In the current study we therefore consider finite-size evaporating droplets in SS-HST, assuming an incompressible liquid surrounded by a compressible gas phase at higher temperature. The initial size, ranging between $10.5\eta$ and $21.5\eta$, is chosen to focus on evaporation enhancement by turbulence~\citep{verwey2020experimental} and to elucidate the effects of the interface deformation. This last aspect is less discussed in literature and, so far, the spherical assumption has been invoked to describe the droplet shape also in fully resolved simulations~\citep{lupo2019immersed,lupo2020direct}. Given the large parameter space which characterizes evaporating flows, here we focus on changing the ratio between the droplet initial diameter and the Kolmogorov scale ($d_0/\eta$), the surface tension and the initial gas temperature. Moreover, in all cases, to study the isolated droplet behaviour, we consider a small initial liquid volume fraction, $\alpha_0=0.14$ $\%$, corresponding to five droplets. The resulting parametric study aims at addressing the following questions:} 
\vspace{0.1 cm}
\begin{itemize}
  \item [a.] What is the level of approximation of the estimated evaporation rates when the gas thermophysical properties are assumed constant?
  \item [b.] What are the effects on the evaporation rate and on the liquid temperature of the droplet size? Moreover, how does the ratio $K/K_0$ (actual evaporation rate in turbulence over the evaporation rate in stagnant conditions) changes when increasing the gas temperature in conditions relevant for combustion applications? 
  \item [c.] How does the interface deformation affect the evaporation rate and does the local interfacial mass flux correlates with changes in the droplet shape?
\end{itemize}
To investigate this complex phenomenon numerically, we propose a new VoF method for evaporating flows in weakly compressible homogeneously sheared turbulence, extending the algorithm in \cite{scapin2020volume}.
The tool addresses the two main issues arising when performing this kind of simulation with more realistic and challenging conditions. First, as already remarked in~\cite{kasbaoui2017algorithm}, numerical simulations in HST are demanding even in single phase since the commonly employed multistep time-integration schemes (e.g., Adams-Bashforth and Runge-Kutta), if employed in their classical formulation, are weakly unstable and, therefore, not adequate for long-time simulations. Next, since the HST computational domain does not possess any outflow boundary, a rigorous description of the two-phase evaporating system requires a compressible formulation that allows the thermodynamic pressure to vary with the state variables. To address the first issue, we present a modified version of the Adams-Bashforth scheme which recovers the analytical solution of Kelvin modes in the limit of the rapid distortion theory (RTD) \citep{maxey1982distortion} and, overall, ensures a stable integration over long times. The second issue is addressed by deriving and presenting a new mathematical formulation for evaporating flows with phase change in the low-Mach limit (weakly compressible formulation) with a detailed numerical implementation. Differently from other approaches available in literature~\citep{wang2019vaporization,ni2021numerical}, this formulation relaxes the assumption of constant thermodynamic pressure and allows its dynamic variation according to the global expansion and contraction of the compressible gas phase, as well as, on the amount of mass flux exchanged at the interface. \\
%
%
This paper is organized as follows. In \S~\ref{sec:gov_eqn} we introduce the mathematical model employed to describe a two-phase evaporating system, adapted to the HST configuration. In \S~\ref{sec:num} the numerical algorithm is presented, with a note on an improved Adams-Bashforth scheme for HST simulations. The results, discussed in \S~\ref{sec:results}, focus on the role played by the specific thermodynamic model used to describe the weakly compressible phase and the effects on the evaporation induced by the variation of the shear-based Reynolds number and the shear-based Weber number. Finally, the main findings and conclusions are summarized in \S~\ref{sec:concl}.

\section{Governing equations}\label{sec:gov_eqn}

\subsection{Mathematical model for weakly compressible evaporating flows}
We consider a system of two immiscible and Newtonian fluids: a single component liquid (phase $1$) and an ideal mixture of an inert gas and vaporized liquid (phase $2$). The two phases are bounded by an infinitesimally small interface, through which mass, momentum and energy are exchanged. The evaporation is driven by the partial pressure of the inert gas in phase $2$. To represent this system, a phase indicator function $H$ is defined at position $\mathbf{x}$ and time $t$ to distinguish between the two phases:
\begin{equation}
H(\mathbf{x},t) = \begin{cases}
				1 \hspace{0.5 cm} \text{if $\mathbf{x} \in V_l$}\mathrm{,} \\
				0 \hspace{0.5 cm} \text{if $\mathbf{x} \in V_g$}\mathrm{;} 
			   \end{cases}
\label{eqn:phase_ind}
\end{equation} 
where $V_l$ and $V_g$ are the domains pertaining to the liquid and gas phases, divided by a zero-thickness interface $\Gamma(t)=V_l\bigcap V_g$. \\ 
\revC{Hereinafter and unless otherwise state, we assume that the liquid is incompressible with constant properties, while the gas is compressible and its properties are allowed to vary with temperature, pressure and composition}. Thus, given the possible variation of the density with the state variables, we consider compressibility, which in this work is treated within the low-Mach approximation~\citep{majda1985derivation}. This allows us to filter acoustic effects, while still retaining potentially large density variations in the bulk region of the compressible phase. Under this assumption, the conservation equations for species, momentum, thermal energy and mass across the interface read as
\begin{equation}
  \rho\dfrac{D\mathbf{u}}{Dt} =-\nabla p+\dfrac{1}{Re}\nabla\cdot\mathbf{\tau}+\dfrac{\mathbf{f}_{\sigma}}{We}\mathrm{,} \label{eqn:mom1}
\end{equation}
\begin{equation}
  \rho_g\dfrac{DY_{l}^{v}}{Dt} = \dfrac{1}{ReSc}\nabla\cdot(\rho_gD_{lg}\nabla Y_{l}^{v}) \label{eqn:vap1}\mathrm{,}
\end{equation}
\begin{align}
  \rho c_p\dfrac{DT}{Dt} &= \dfrac{1}{RePr}\nabla\cdot(k\nabla T) + \left(\Pi_{p,1}\dfrac{dp_{th}}{dt}+\dfrac{\rho_gD_{lg}}{ReSc}\sum_{j=1}^{2}\nabla h_j\cdot\nabla Y_j\right)(1-H)\label{eqn:tmp1} \\ 
&- \dfrac{(\dot{m}_{\Gamma}\delta_{\Gamma})}{Ste}\mathrm{,} \nonumber
\end{align}
\vspace{-0.45 cm}
\begin{equation}
  \dot{m}_{\Gamma} = \dfrac{1}{ReSc}\dfrac{\rho_{g,\Gamma}D_{lg,\Gamma}}{1-Y_{l,\Gamma}^v}\nabla_{\Gamma} Y_{l}^{v}\cdot\mathbf{n}_{\Gamma}\label{eqn:int_mfx}\mathrm{.}
\end{equation}
\nsc{Here, $\mathbf{u}$ is the fluid velocity assumed to be continuous in the two phases, $p$ is the hydrodynamic pressure, $T$ the temperature, $h$ the enthalpy (with $\nabla h=\nabla(c_pT)$), $Y_{l}^v$ the mass fraction of the vaporized liquid in the inert gas and $\dot{m}_{\Gamma}$ is the interfacial mass flux.} In~\eqref{eqn:mom1}, $\tau$ is the viscous stress tensor for compressible Newtonian flows and $\mathbf{f}_{\sigma}=\kappa_{\Gamma}\delta_{\Gamma}$ with $\kappa_{\Gamma}$ the interfacial curvature. The generic thermophysical property $\xi$ (density $\rho$, dynamic viscosity $\mu$, thermal conductivity $k$ or specific heat capacity $c_p$) is computed with an arithmetic average, i.e. $\xi=1+(\lambda_{\xi}-1)H$ where $\lambda_{\xi}=\xi_l/\xi_{g,r}$. Since $\xi_l$ is kept constant and uniform no further modelling is needed, while the generic gas property $\xi_{g}$ is computed with appropriate equation of states. The gas density $\rho_g$ is computed with the ideal gas law and the liquid diffusion coefficient with the Wilke-Lee correlation~\citep{reid1987properties}. More details on how the remaining gas thermophysical properties are evaluated are given in Appendix~\ref{sec:eqn_state}. Unless otherwise stated, all the property ratios $\lambda_{\xi}$ are computed with respect to the reference gas property $\xi_{g,r}$, taken as the initial value. \\ 
In equations~\eqref{eqn:mom1},~\eqref{eqn:vap1},\eqref{eqn:int_mfx} and~\eqref{eqn:tmp1} different dimensionless parameters appear. By introducing a reference velocity $u_{r}$ and reference length $l_{r}$, we define the Reynolds number $Re=\rho_{g,r}u_{r}l_{r}/\mu_{g,r}$, the Weber number $We=\rho_{g,r}u_{r}^2l_{r}/\sigma$, with $\sigma$ the surface tension; $Sc=\mu_{g,r}/(\rho_{g,r}D_{lg,r})$ and $Pr=\mu_{g,r}c_{pg,r}/k_{g,r}$ are the Schmidt and the Prandlt numbers. Note that the temperature equation~\eqref{eqn:tmp1} requires the definition of the Stefan number $Ste=c_{pg,r}T_{g,0}/\Delta h_{lv}$, where $T_{g,0}$ is the initial gas temperature and $\Delta h_{lv}$ is the latent heat and of the dimensionless group $\Pi_{p,1}=R_u/(c_{pg,r}M_g)$ where $M_g$ is the molar mass of the gas phase and $R_u$ the universal gas constant. To form a close set of equations, two additional equations are needed, one for the velocity divergence and one for the thermodynamic pressure $p_{th}$, i.e.
\begin{equation}
  \nabla\cdot\mathbf{u} = f_{\Gamma}(\mathbf{x}_{\Gamma},t) + \left[\dfrac{f_{Y}(\mathbf{x},t) + f_T(\mathbf{x},t)}{p_{th}} - \left(1-\dfrac{\Pi_{p,1}}{c_p\overline{M}_{m,av}}\right)\dfrac{1}{p_{th}}\dfrac{dp_{th}}{dt}\right](1-H)
  \label{eqn:veldiv1}\mathrm{,}
\end{equation}
\begin{equation}
  \dfrac{1}{p_{th}}\dfrac{dp_{th}}{dt}\int_{V_g}\left(1-\dfrac{\Pi_{p,1}}{c_p\overline{M}_{m,av}}\right)dV_g = \int_V\left[f_{\Gamma}(\mathbf{x}_{\Gamma},t)+\dfrac{f_T(\mathbf{x},t)+f_Y(\mathbf{x},t)}{p_{th}}(1-H)\right] dV 
  \label{eqn:pth1}\mathrm{.}
\end{equation}
In equations~\eqref{eqn:veldiv1} and~\eqref{eqn:pth1} the functions $f_{\Gamma}(\mathbf{x}_{\Gamma},t)$, $f_{Y}(\mathbf{x},t)$ and $f_{T}(\mathbf{x},t)$ represent the different contributions to the total velocity divergence from the phase change ($f_{\Gamma}$) and the change of the gas density either due to composition ($f_Y$) or to temperature ($f_T$),
\begin{subequations}
\begin{align}
  &f_{\Gamma}(\mathbf{x}_{\Gamma},t) = \dot{m}_{\Gamma}\left(\dfrac{1}{\rho_{g,\Gamma}}-\dfrac{1}{\lambda_{\rho}}\right)\delta_{\Gamma}\mathrm{,} \\
  &f_{Y}(\mathbf{x},t) = \dfrac{1}{ReSc}\dfrac{\overline{M}_{m,av}}{\rho_g}\left(\dfrac{1}{\lambda_M}-1\right)\nabla\cdot(\rho_gD_{lg}\nabla Y_{l}^{v})\mathrm{,} \\
	&f_{T}(\mathbf{x},t) =\dfrac{1}{Re}\dfrac{\Pi_{p,1}}{c_p\overline{M}_{m,av}}\left[\dfrac{1}{Pr}\nabla\cdot(k\nabla T)+\dfrac{\rho_gD_{lg}}{Sc}\sum_{j=1}^2\nabla h_j\cdot\nabla Y_j\right]\mathrm{,}
\end{align}
  \label{f_aux}
\end{subequations}
\hspace{-0.25 cm}where $\lambda_M=M_l/M_g$ is the molar mass ratio. The complete derivation of equations~\eqref{eqn:veldiv1},~\eqref{eqn:pth1} and of relations~\eqref{f_aux} is provided in Appendix~\ref{sec:vel_div}. 

\subsection{Governing equations for the HST configuration}
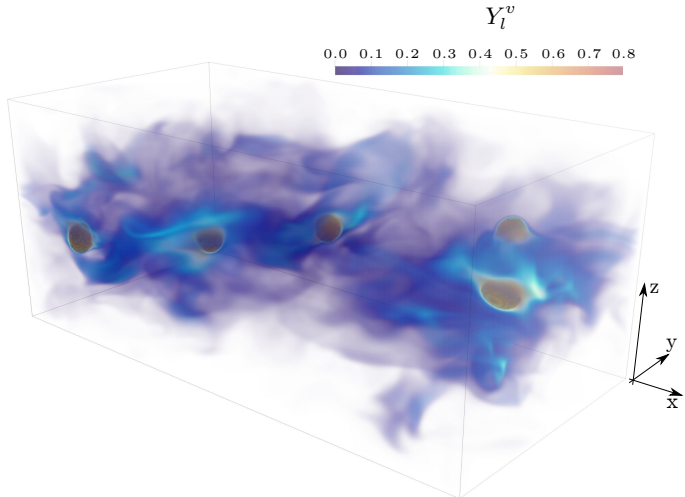
\begin{figure}
    \centering
    \includegraphics[height=6.185 cm,width=9.5 cm]{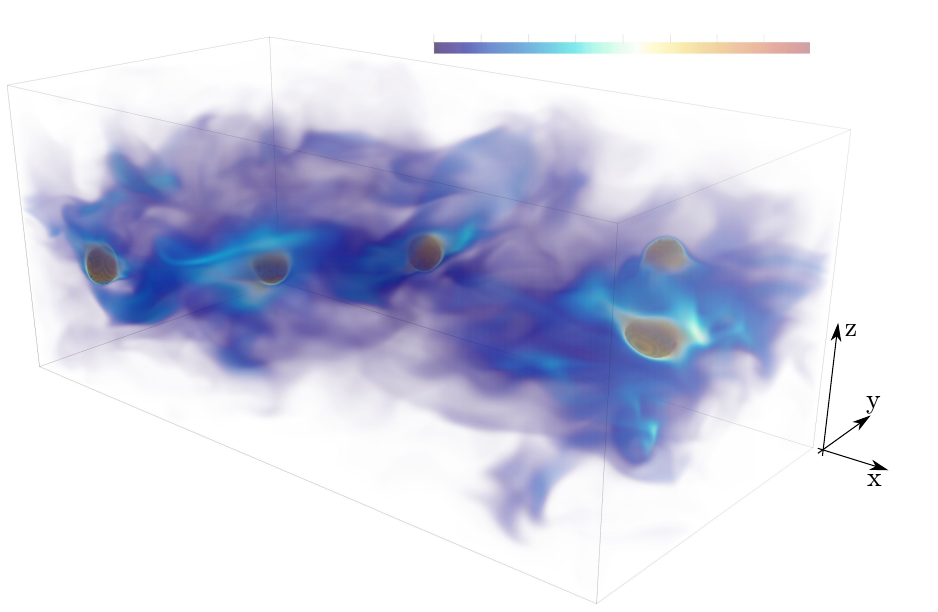}
    \put(-88.5,180){\small{$Y_{l}^{v}$}}
    \put(-149.5,167){\tiny{$0.0$}}
    \put(-136.0,167){\tiny{$0.1$}}
    \put(-122.5,167){\tiny{$0.2$}}
    \put(-108.5,167){\tiny{$0.3$}}
    \put(-95.00,167){\tiny{$0.4$}}
    \put(-81.50,167){\tiny{$0.5$}}
    \put(-67.50,167){\tiny{$0.6$}}
    \put(-54.00,167){\tiny{$0.7$}}
    \put(-41.00,167){\tiny{$0.8$}}
    \caption{Sketch of the computational box: the rendering represents the volume contour of the vapour mass fraction for the case at $Re_{\mathcal{S}}=6700$, $We_{\mathcal{S}}=0.1$ and $T_{g,0}/T_c=1.5$.}
    \label{fig:box_shear}
\end{figure}

Equations~\eqref{eqn:mom1},~\eqref{eqn:vap1},~\eqref{eqn:int_mfx},~\eqref{eqn:tmp1},~\eqref{eqn:veldiv1} and~\eqref{eqn:pth1} are solved in a periodic box assuming an imposed uniform mean shear $\mathcal{S}$, as depicted in figure~\ref{fig:box_shear}. In a shear-periodic domain the streamwise $x$ and spanwise $y$ directions are periodic, whereas the so-called shear-periodic condition applies in the $z$ direction, which reads for the generic scalar quantity $g$ as
\revA{
\begin{equation}
  \begin{aligned}
    &g(x+l_x,y,z) = g(x,y,z)\mathrm{,} \\
    &g(x,y+l_y,z) = g(x-\mathcal{S}tl_y,y,z)\mathrm{,} \\
    &g(x,y,z+l_z) = g(x,y,z)\mathrm{.}
  \end{aligned}
\end{equation}
}

The presence of a mean velocity, i.e., $\mathcal{S}z$, suggests to decompose the velocity field $\mathbf{u}$ into a mean and a fluctuating component $\mathbf{u}'$,
\begin{equation}
	\mathbf{u}=\mathbf{u}'+\mathcal{S}z\mathbf{e}_x\mathrm{,}
  \label{eqn:decomp}
\end{equation}
where $\mathbf{e}_x=(1,0,0)$. 
Given the decomposition~\eqref{eqn:decomp}, the momentum equation~\eqref{eqn:mom1} is written in terms of $\mathbf{u}'$:
\begin{equation}
	\rho\left(\dfrac{D\mathbf{u}'}{Dt} + \mathcal{S}z\dfrac{\partial\mathbf{u}'}{\partial x} + \mathcal{S}w'\mathbf{e}_x\right) = -\nabla p + \dfrac{1}{Re_{\mathcal{S}}}\left[\nabla\cdot\tau+\mathcal{S}\left(\dfrac{\partial\mu}{\partial z}\mathbf{e}_x+\dfrac{\partial\mu}{\partial x}\mathbf{e}_z\right)\right]+\dfrac{\mathbf{f}_{\sigma}}{We_{\mathcal{S}}}\mathrm{,}
  \label{eqn:mom2}
\end{equation}
where $\mathbf{e}_{z}=(0,0,1)$. Three new terms appear: the first, $\mathcal{S}z\left(\partial\mathbf{u}'/\partial x\right)$, is the convection of the velocity fluctuations by the mean shear; the second, $\mathcal{S}w'\mathbf{e}_x$, represents the production of streamwise momentum caused by the fluid parcel transport in the normal direction owing to the mean shear; the third and last term, $\mathcal{S}\left(\partial\mu/\partial z\mathbf{e}_x+\partial\mu/\partial x\mathbf{e}_z\right)$, represents the viscous dissipation due to the mean shear  in the case of a fluid with variable viscosity. Note that $Re_{\mathcal{S}}$ and $We_{\mathcal{S}}$ are the shear-based Reynolds and Weber numbers, computed taking $u_{r}=\mathcal{S}l_r$. \par
The same decomposition~\eqref{eqn:decomp} is applied to the mass fraction and temperature equations, giving
\begin{equation}
	\rho_g\left(\dfrac{DY_{l}^{v}}{Dt}+\mathcal{S}z\dfrac{\partial Y_{l}^{v}}{\partial x}\right) = \dfrac{1}{Re_{\mathcal{S}}Sc}\nabla\cdot(\rho_gD_{lg}\nabla Y_{l}^{v})\mathrm{,}  \label{eqn:vap3} 
\end{equation}
\begin{align}
	\rho c_p\left(\dfrac{DT}{Dt} + \mathcal{S}z\dfrac{\partial T}{\partial x}\right) &= \dfrac{1}{Re_{\mathcal{S}}Pr}\nabla\cdot(k\nabla T) \\ &+\left(\Pi_{p,1}\dfrac{dp_{th}}{dt}+\dfrac{\rho_gD_{lg}}{Re_{\mathcal{S}}Sc}\sum_{j=1}^{2}\nabla h_j\cdot\nabla Y_j\right)(1-H)-\dfrac{\dot{m}_{\Gamma}\delta_{\Gamma}}{Ste}\mathrm{.}
  \label{eqn:tmp3}
\end{align}
In equations~\eqref{eqn:tmp3} and~\eqref{eqn:vap3} the two new terms, $\mathcal{S}z\left(\partial T/\partial x\right)$ and $\mathcal{S}z\left(\partial Y_{l}^{v}/\partial x\right)$, represent the convection of the temperature/vapour mass fraction by the mean shear.

\section{Methodology}\label{sec:num}

\subsection{Numerical method for low-Mach HST simulations with phase change}
The governing equations are solved on a uniform Cartesian grid of equal spacing $\Delta x=\Delta y=\Delta z$, with a staggered arrangement for the velocity while the remaining scalar fields are defined at the cell centres. The convection terms of the governing equations are discretized with the QUICK scheme~\citep{leonard1979stable}, while central schemes are employed for the diffusive terms. The numerical method for phase change in the incompressible limit is presented in~\citep{scapin2020volume}, while the details of the weakly compressible two-phase code are reported in~\citep{dalla2020interface}. 
Both implementations are based on the solver CaNS~\citep{costa2018fft}, extended in this work to handle shear-periodic boundary conditions. In this section we therefore only describe the main modifications needed to handle weakly compressible phase-change processes in homogeneous shear turbulence (HST). The validation of the present algorithm against \revA{three} benchmarks is provided in Appendix~\ref{sec:val_lm}.
 
\subsubsection{Dispersed phase}
The first step is the interface reconstruction and subsequent advection, handled in a fully Eulerian manner using an algebraic VoF method, i.e.,\ the MTHINC by~\cite{ii2012interface,rosti2019numerical}. We start from the topological equation for the phase indicator function~\eqref{eqn:phase_ind}:
\begin{equation}
	\dfrac{\partial H}{\partial t} + \mathbf{u}_{\Gamma}\cdot\nabla H = 0\mathrm{,}
  \label{eqn:phi}
\end{equation}
where $\mathbf{u}_{\Gamma}$ is the interface velocity, taken as the sum of the extended liquid velocity $\mathbf{u}_l^{ext}$ and the contribution due to the interfacial mass flux $\dot{m}_{\Gamma}\mathbf{n}_{\Gamma}/\rho_l$; see~\cite{scapin2020volume} for more details. Note that since $\mathbf{u}_{l}^{ext}$ and $\mathbf{u}_{\Gamma}$ are derived from the one-fluid velocity $\mathbf{u}$, the decomposition~\eqref{eqn:decomp} applies directly to the interface velocity. \par 
Equation~\eqref{eqn:phi} is then rewritten in terms of the volume fraction, $\Phi$, defined as the volumetric average of $H$ over a discrete computation cell of volume $V_c=\Delta x\Delta y\Delta z$. Employing the decomposition~\eqref{eqn:decomp} in the colour function advection equation~\eqref{eqn:phi} yields
\begin{equation}
	\dfrac{\partial\Phi}{\partial t} + \nabla\cdot(H^{ht}\mathbf{u}'_{\Gamma}) + \mathcal{S}z\dfrac{\partial H^{ht}}{\partial x} = \Phi\nabla\cdot\mathbf{u}'_{\Gamma}\mathrm{,}
  \label{eqn:phi1}
\end{equation}
where $H^{ht}$ represents the hyperbolic tangent function approximating the indicator function $H$. From equation~\eqref{eqn:phi1}, we see that an additional term is present, $\mathcal{S}z(\partial H^{ht}/\partial x)$, which represents the convection of volume fraction by $\mathcal{S}$. \par
Equation~\eqref{eqn:phi1} is solved in three sub-steps. First, it is advanced by $\Delta t^{n+1}$ to the new time step  omitting the convection by the mean shear and the right-hand side term. By employing the classical directional splitting, we obtain a provisional $\tilde{\Phi}$, see~\cite{ii2012interface} for more details. Next, the convection by mean shear is included
\begin{equation}
	\dfrac{\partial\tilde{\Phi}}{\partial t}+\mathcal{S}z\dfrac{\partial \tilde{H}^{ht}}{\partial x} = 0\mathrm{.}
  \label{eqn:phi2}
\end{equation}
Note that the time derivative in equation~\eqref{eqn:phi2} is computed explicitly for $\tilde{\Phi}$ while the convective term contains the phase indicator function $H^{ht}$. Since the latter is a function of the former and the mathematical form of this dependency varies according to the type of interface reconstruction method, it is possible to express the convective term as a function of $\Phi$. Nevertheless, this would modify the advection velocity in~\eqref{eqn:phi2} adding a spatial-dependent term along the mean shear direction (i.e., $x$), making more elaborated and complex the application of the method of characteristics. For this reason, we prefer to
compute this extra term by an additional directional splitting along $x$, with $\mathcal{S}z$ as advection velocity. \\
Finally, the divergence is corrected. This consists in adding, after the four directional splittings, a correction term proportional to the discrete velocity divergence, i.e.
\begin{equation}
	\Phi_{i,j,k}^{n+1} = \tilde{\Phi}_{i,j,k} - \Delta t^{n+1}F_{i,j,k}^n+\Delta t^{n+1}\Phi_{i,j,k}^{n+1}(\nabla\cdot\mathbf{u}_{\Gamma})_{i,j,k}^n\mathrm{,}
\end{equation}
where $\Phi_{i,j,k}^{n+1}$ is the volume fraction resulting from the directional splitting procedure, $F_{i,j,k}^n$ represents the correction used in the previous directional splitting steps, and the last term is the volume correction  which ensures that the interface velocity divergence is employed to update 
$\Phi_{i,j,k}^{n+1}$~\citep{scapin2020volume}. The thermodynamic properties ($\rho$, $\mu$, $k$ and $c_p$) are then updated using the new value $\Phi^{n+1}$. 

\subsubsection{Vapour mass fraction equation}\label{sec:vap_mass}
The conservation of the vapour mass fraction, see equation~\eqref{eqn:vap3}, is solved only in the gas domain (i.e., $V_g$), assuming saturation conditions at the interface $\Gamma$. 
\nsc{In other words, the value $Y_{l}^{v}=Y_{l,\Gamma}^{v}$, which is a function of the thermodynamic pressure and temperature, is imposed at the interface $\Gamma$ as a Dirichlet boundary condition. To compute $Y_{l,\Gamma}^{v}$, we employ Span-Wagner equation of state (see equations~\eqref{eqn:sp_wg_1} and~\eqref{eqn:sp_wg_2} in Appendix~\ref{sec:eqn_state}). Equation~\eqref{eqn:vap3} is advanced with the first-Euler method neglecting the mean shear contribution.} This yields a provisional vapour mass fraction field $\tilde{Y}_{v,l}^{n+1}$ :
\begin{equation}
	\rho_g^{n+1}\dfrac{\tilde{Y}_{l}^{v,n+1}-Y_{l}^{v,n}}{\Delta t^{n+1}} = -\rho_g^{n+1}(\mathbf{u}\cdot\nabla Y_{l}^v)^n + \dfrac{1}{Re_{\mathcal{S}}Sc}\nabla\cdot(\rho_g^{n+1}D_{lg}^{n+1}\nabla Y_{l}^{v,n})\mathrm{.}
  \label{eqn:vap3_disc}
\end{equation}
The calculation of the gradient in the convective part of~\eqref{eqn:vap3_disc} is performed as detailed in~\cite{scapin2020volume}, while some modifications are required for the linear term as it contains the diffusion coefficient $D_{lg}$ and the gas density $\rho_g$, which, in general, vary with the thermodynamic pressure, temperature and composition. Since the procedure can be performed dimension by dimension, we present the discretization along $x$ as example, the same approach being repeated for the other two directions,
\begin{equation}
  \dfrac{\partial}{\partial x}\left(\rho_gD_{lg}\dfrac{\partial Y_{l}^{v}}{\partial x}\right)_{i} = \dfrac{1}{\Delta x}\left((\rho_gD_{lg})_{i+1/2}\left.\dfrac{\partial Y_{l}^{v}}{\partial x}\right|_{i+1/2}-\left.(\rho_gD_{lg})_{i-1/2}\dfrac{\partial Y_{l}^{v}}{\partial x}\right|_{i-1/2}\right)_i\mathrm{.}
\end{equation}
The evaluation of the gradients $\partial Y_{l}^{v}/\partial x_{i\pm 1/2}$ is performed on an irregular grid, employing one-sided finite difference equations for the cell cut from the interface or central scheme for cells away from the interface \cite[see][for details]{scapin2020volume}. Next, the coefficients $(\rho_gD_{lg})_{i\pm 1/2}$ are obtained as the arithmetic mean,
\begin{equation}
	(\rho_gD_{lg})_{i\pm 1/2} = 0.5[(\rho_gD_{lg})_i+(\rho_gD_{lg})^G]\mathrm{.}
\end{equation}
If the cell $i$ and its neighbours ($i\pm 1$) are not cut by the interface, $\rho_gD_{lg}^G$ is set equal to $(\rho_gD_{lg})_{i\pm 1}$. Otherwise, $(\rho_gD_{lg})^G$ is evaluated as
\begin{equation}
	(\rho_gD_{lg})^G = \dfrac{(\rho_gD_{lg})_{\Gamma}+(\theta-1)(\rho_gD_{lg})_{i\pm 1}}{\theta}\mathrm{,}
  \label{eqn:alphag}
\end{equation}
where $\theta$ represents the sub-cell resolution computed from the level-set function, reconstructed from the volume of fluid field. 
Since eq.~\eqref{eqn:alphag} poorly behaves for small values of $\theta$, we set $(\rho_gD_{lg})^G=(\rho_gD_{lg})_i$ when $\theta\leq 1/4$. Note that equation~\eqref{eqn:alphag} requires the value of the coefficient $(\rho_gD_{lg})$ at the interface location, i.e. $(\rho_gD_{lg})_{\Gamma}$. 
These are evaluated with the corresponding equations of state using the temperature and the vapour composition at the interface location. It is worth 
mentioning that we cannot access directly the liquid and gas temperatures, separately, since the temperature equation is solved over the whole domain, irrespective of the interface location. Therefore, to avoid problems of artificial heating (especially when the difference between the gas and the liquid density is high), we locally reconstruct the gas temperature in $V_l$ and the liquid temperature in $V_g$ relying on a simple constant extrapolation of $T$. The resulting two fields, $T_g$ and $T_l$, defined in a few grid cells around the interface, are then used to update the thermophysical properties, see Appendix~\ref{sec:eqn_state} for more details. \par
Finally, the mean shear contribution is included using the method of characteristic over a time $\Delta t^{n+1}$:
\begin{equation}
  \dfrac{\partial Y_{l}^{v}}{\partial t}+\mathcal{S}z\dfrac{\partial Y_{l}^{v}}{\partial x} = 0\mathrm{.}
  \label{eqn:yvap_rem}
\end{equation}
Equation~\eqref{eqn:yvap_rem} can be conveniently rewritten in more compact form as in~\cite{gerz1989direct,kasbaoui2017algorithm}:
\begin{equation}
	Y_{l}^{v,n+1} = \tilde{Y}_{l}^{v,n+1}(\mathbf{x}-\Delta t^{n+1}\mathcal{S}z\mathbf{e}_x)\mathrm{.}
  \label{eqn:yv}
\end{equation}
Equation~\eqref{eqn:yv} is solved using a Fourier interpolation~\citep{tanaka2017effect}, which ensures a spectral accuracy provided that $\Delta t^{n+1}$ is chosen lower or equal than $(\mathcal{S}N_z)^{-1}$ where $N_z$ is the number of grid points along the $z$ direction (see section~\ref{sec:time_step} for more details).

\subsubsection{Calculation of the interfacial mass flux}\label{sec:mfx}
The interfacial mass flux~\eqref{eqn:int_mfx} is computed only in the gas region by projecting the interfacial gradient along the normal direction and adopting a dimension by dimension approach. The interfacial vapour mass fraction $Y_{l}^{v}$, gas density $\rho_{g}$ and diffusion coefficient $D_{lg}$ should be estimated at the interface location. Similarly to the case of the vapour mass fraction described in section~\ref{sec:vap_mass}, all these quantities depend on the local gas and liquid temperatures and on the thermodynamic pressure. Therefore, we first estimate the interfacial liquid and gas temperature in those grid cells cut by the interface. Depending on the interface position, this is done in each direction independently, providing different estimates for the gas and liquid temperature for the three coordinates, $T_{p,\Gamma}^x$, $T_{p,\Gamma}^y$ and $T_{p,\Gamma}^z$ where the subscript $p$ stands for the gas and liquid phase. The resulting values are averaged using the local normal, i.e.:
\begin{equation}
	T_{p,\Gamma} = T_{p,\Gamma}^xn_x^2 + T_{p,\Gamma}^yn_y^2 + T_{p,\Gamma}^zn_z^2\mathrm{.}
\end{equation}
Once $T_{p,\Gamma}$ is known, the values of $\rho_{g,\Gamma}$, $D_{lg,\Gamma}$ and $Y_{l,\Gamma}^v$ are computed using the corresponding equations of state, see Appendix~\ref{sec:eqn_state}. \nsc{Note that by employing the procedure here explained,} the mass flux $\dot{m}_{\Gamma}$ is available only on the grid nodes pertaining the gas region. Nevertheless, as equation~\eqref{eqn:veldiv1} suggests, the values of $\dot{m}_{\Gamma}$ are needed also in some grid points inside the liquid region. Accordingly, $\dot{m}_{\Gamma}$ is extrapolated over a narrow band at the interface to populate all cells where $|\nabla \Phi|_{i,j,k}\neq 0$.

\subsubsection{Temperature equation}\label{sec:tmp}
The temperature equation~\eqref{eqn:tmp1} is solved using a whole domain approach as in~\cite{scapin2020volume}, with additional care paid to the time discretization. \nsc{The adopted approach follows the one proposed in~\cite{gerz1989direct} and has been improved here to enhance the numerical stability and to include the additional source terms due to the gas compressibility, phase change and enthalpy diffusion.}
First, a prediction temperature field $\tilde{T}$ is computed using the Adams-Bashforth method:
\begin{equation}
  \dfrac{\tilde{T}^{n+1}-T^n}{\Delta t^{n+1}} = \left(1+\dfrac{1}{2}\dfrac{\Delta t^{n+1}}{\Delta t^n}\right)RT^n-\left(\dfrac{1}{2}\dfrac{\Delta t^{n+1}}{\Delta t^n}\right)RT^{n-1}\mathrm{,}
  \label{eqn:disc_tmp}
\end{equation}
where the terms $RT^n$ and $RT^{n-1}$ include all the convective and diffusive terms at the current, $n$, and old time level, $n-1$. 
The term $RT^{n-1}$ is first shear mapped to the new time level $n+1$ and this step has proved to be crucial for the numerical stability of the Adams-Bashforth scheme (see Appendix~\ref{sec:imp_ab2}). \par
Next, the temperature field is shear mapped to the new time level $n+1$ by employing the same spectral interpolation described for the vapour mass fraction equation,
\begin{equation}
  T^{n+1,*} = \tilde{T}(\mathbf{x}-\mathcal{S}z\Delta t^{n+1}\mathbf{e}_x)\mathrm{.}
  \label{eqn:remap_tmp}
\end{equation}
Finally, the source terms in~\eqref{eqn:tmp1} are included using the first-order Euler scheme:
\begin{align}
	\rho^{n+1}c_p^{n+1}\dfrac{T^{n+1}-T^{n+1,*}}{\Delta t^{n+1}} &= \,\, \Pi_{p,1}\left(\dfrac{dp_{th}}{dt}\right)^{ext}(1-\Phi^{n+1})\mathrm{,} \label{eqn:tmp_disc}\\ & \nonumber
	  +\dfrac{1}{Re_{\mathcal{S}}Sc}\left(\rho_gD_{lg}\sum_{j=1}^2\nabla Y_j\cdot\nabla h_j^{*}\right)^{n+1}(1-\Phi^{n+1})\mathrm{,} \\ & \nonumber
          - \dfrac{(\dot{m}_{\Gamma}|\nabla \Phi|)^{n+1}}{Ste}\mathrm{,}
\end{align}
where $(dp_{th}/dt)^{ext}$ represents the linear extrapolation in time of the derivative of the thermodynamic pressure. It is important to note that the second and third terms in equation~\eqref{eqn:tmp_disc}, which include variables shear mapped to the new time level, (i.e., $\Phi^{n+1}$ and $Y_{l,j}^{v,n+1}$) should be computed after~\eqref{eqn:remap_tmp}.

\subsubsection{Thermal divergence and thermodynamic pressure}\label{subsec:divth_pth}
The calculation of the velocity divergence is performed simply by discretizating the right-hand side of equation~\eqref{eqn:veldiv1} node by node as done in~\cite{dalla2020interface} for the case of weakly compressible two-phase solvers without phase change. The calculation of the thermodynamic pressure requires, however, more care. 
\nsc{In  the low-Mach framework, the role of $p_{th}$ is to ensure mass conservation of the compressible phase at the discrete level, since it enters the calculation of the gas density. This cannot be fulfilled simply by the advection of the color-function, which is designed to satisfy the volume conservation of the incompressible liquid, or by the pressure-correction step through the imposition of the divergence constrain~\eqref{eqn:veldiv1} on $\mathbf{u}$. In fact, these ensure only the overall volume conservation of the closed and isochoric system under consideration. Therefore, we compute $p_{th}$  by integrating the equation for the gas density (equation~\ref{gas_density} in Appendix~\ref{sec:eqn_state}) over the computation domain occupied by the gas phase $V_g$ \citep{demou2019low,dalla2020interface},}
\begin{equation}
	p_{th}^{n+1} = \dfrac{G_{g}^{n+1}\Pi_{p,2}}{\displaystyle{\int_{V}\dfrac{\overline{M}_{m,av}^{n+1}}{T^{n+1}}(1-\Phi^{n+1})\,dx\,dy\,dz}}\mathrm{.}
\end{equation}
The total mass of the gas $G_g$, used above, varies in time according to the following relation, derived from the material balance across the droplet surface,
\begin{equation}
	\dfrac{dG_g}{dt} = \dot{m}_{t,\Gamma} = \int_{V}\dot{m}_{\Gamma}^{n+1}|\nabla \Phi^{n+1}|\,dx\,dy\,dz\mathrm{.}
  \label{eqn:cal_mg1}
\end{equation}
Once the volume integral in equation~\eqref{eqn:cal_mg1} is computed, the gas mass at the new time level is computed from the time integration of~\eqref{eqn:cal_mg1},
\begin{equation}
	G_{g}^{n+1} = G_{g}^{n} + \int_{t^n}^{t^{n+1}}\dot{m}_{t,\Gamma}\,dt\mathrm{.}
  \label{eqn:cal_mg2}
\end{equation}
To evaluate numerically the integral in~\eqref{eqn:cal_mg2}, we employ a trapezoidal quadrature. With this approach, 
we impose correctly the conservation of the compressible phase at the discrete level.

\subsubsection{Momentum equation and pressure correction method}\label{subsec:momeqn}
Once the thermodynamic divergence is computed, the momentum equation~\eqref{eqn:mom1} is solved with a standard pressure correction method, reported below in semi-discrete form,
\begin{align}
    \mathbf{RU}^{n-1} = \mathbf{\widetilde{RU}}^{n-1}(\mathbf{x}-\Delta t^{n}\mathcal{S}z\mathbf{e}_x)\mathrm{,}\\
    \rho^{n+1}\left(\dfrac{\mathbf{u}^{\star\star}-\mathbf{u}^{n}}{\Delta t^{n+1}}\right) = \left(1+\dfrac{1}{2}\dfrac{\Delta t^{n+1}}{\Delta t^n}\right)\mathbf{RU}^{n}-\left(\dfrac{1}{2}\dfrac{\Delta t^{n+1}}{\Delta t^n}\right)\mathbf{RU}^{n-1}\mathrm{,}\label{eqn:u_predic} \\
	\mathbf{u}^{\star} = \mathbf{u}^{\star\star}(\mathbf{x}-\Delta t^{n+1}\mathcal{S}z\mathbf{e}_x) + \dfrac{\Delta t^{n+1}}{\rho^{n+1}}\dfrac{\kappa_{\Gamma}^{n+1}\nabla \Phi^{n+1}}{We_{\mathcal{S}}}\mathrm{,} \label{eqn:surf} \\
    \nabla^2p^{n+1} = \nabla\cdot\left[\left(1-\dfrac{\rho_0^{n+1}}{\rho^{n+1}}\right)\nabla\hat{p}\right]+
	\dfrac{\rho_0^{n+1}}{\Delta t^{n+1}}\left(\nabla\cdot\mathbf{u}^\star-\nabla\cdot\mathbf{u}^{n+1}\right)\mathrm{,}\label{eqn:lap_p}\\
	\mathbf{u}^{n+1} = \mathbf{u}^\star-\Delta t^{n+1}\left[\dfrac{1}{\rho_0^{n+1}}\nabla p^{n+1}+\left(\dfrac{1}{\rho^{n+1}}-\dfrac{1}{\rho_0^{n+1}}\right)\nabla\hat{p}\right]\label{eqn:corr}\mathrm{,}
\end{align}
	where $\mathbf{RU}^n$ and $\mathbf{RU}^{n-1}$ in equation~\eqref{eqn:u_predic} include the convective and diffusive terms computed at the current and previous time level, neglecting the surface tension force and the mean shear contribution. As done for the temperature equation and explained in detail in the Appendix~\ref{sec:imp_ab2}, it is very important for the stability and the accuracy of the time integration scheme that the term $\mathbf{RU}^{n-1}$ is shear mapped to the current time level $n$. \nsc{The intermediate velocity $\mathbf{u}^{\star\star}$ is shear mapped to the new time level $n+1$, similarly to what done for the temperature and vapour mass fraction, and updated with the contribution from the surface tension \citep{rosti2019droplets}.} \par
Finally, the pressure equation~\eqref{eqn:lap_p} is solved with a time-splitting technique which allows to transform the variable Poisson equation into a constant coefficient one~\citep{dodd2014fast}. 
Note that $\rho_0^{n+1}$ and $\hat{p}$ in~\eqref{eqn:lap_p} and~\eqref{eqn:corr} are the minimum density over the entire computational domain and the extrapolated hydrodynamic pressure $\hat{p}=(1+(\Delta t^{n+1}/\Delta t^{n}))p^n-(\Delta t^{n+1}/\Delta t^{n})p^{n-1}$. As already discussed in~\cite{motheau2016high,dalla2020interface}, \nsc{the use of the pressure splitting to solve the variable density Poisson equation in a low-Mach framework is suitable when the temperature ratio between the two phases is below $2-3$, which is the case of the current study.} Finally, it is worth mentioning that the presence of the shear, if left untreated, would make the use of the eigenexpansion method to solve~\eqref{eqn:lap_p} not possible, since one direction is not periodic. For this reasons, in order to still benefit from FFT-based solvers, equation~\eqref{eqn:lap_p} is solved in a coordinate system moving with the mean shear for which, triperiodic boundary conditions can be applied. The solution is then transformed back to the shear-periodic domain, as detailed in~\cite{tanaka2017effect,rosti2019droplets}. 

\subsection{Time-step restriction}\label{sec:time_step}
The time step $\Delta t^{n+1}$ is estimated from the stability constraints of the overall system:
%
\begin{equation}
	\Delta t^{n+1}=\min\left[C_{\Delta t_c}\Delta t_c^{n+1},C_{\Delta t_d}\min\left(\Delta t_{\mu},\Delta t_{\sigma},\Delta t_m,\Delta t_k\right)^{n+1},C_{\Delta t_\mathcal{S}}\dfrac{\Delta z}{\mathcal{S}l_z}\right]\mathrm{,}
	\label{eqn:max_dt}
\end{equation}
%
where $\Delta t_c$, $\Delta t_{\sigma}$, $\Delta t_{\mu}$, $\Delta t_m$ and $\Delta t_k$ are the maximum allowable time steps due to convection, surface tension, momentum, thermal energy and mass diffusion. These can be determined as suggested in~\cite{scapin2020volume,dalla2020interface}:
%
\begin{equation}
	\begin{aligned}
	\Delta t_c & = \dfrac{\Delta x}{|u_{\max}|+|v_{\max}|+|w_{\max}|}\mathrm{,} \hspace{1.35 cm} \Delta t_{\mu}=\min\left(\min_{V_g}\left\{\dfrac{\mu_g}{\rho_g}\right\},\dfrac{\lambda_{\mu}}{\lambda_{\rho}}\right)\dfrac{\Delta x^2Re_{\mathcal{S}}}{6}\mathrm{,} \\
		\Delta t_{\sigma}&=\sqrt{\dfrac{We_{\mathcal{S}}(\min_{V_g}\{\rho_g\}+\lambda_{\rho})\Delta x^3}{4\pi}}\mathrm{,} \hspace{0.5 cm}  \Delta t_{m}=\min_{V_g}\left\{D_{lg}\right\}\dfrac{(\theta_m\Delta x)^2Sc}{6}\mathrm{,} \\
\Delta t_{k}&=\min\left(\min_{V_g}\left\{\dfrac{k_g}{\rho_gc_{pg}}\right\},\dfrac{\lambda_{k}}{\lambda_{\rho}\lambda_{c_p}}\right)\dfrac{\Delta x^2Pr}{6}\mathrm{,}
	\end{aligned}
	\label{eqn:diff_con_t}
\end{equation}
%
where $|u_{i,\max}|$ is an estimate of the maximum value of the $i^{th}$ component of the flow velocity, $\theta_m=0.25$ and $\max_{V_g}\{\xi_g\}$ and $\min_{V_g}\{\xi_g\}$ denote the maximum and minimum over the computational domain, $V_g$, of the generic thermophysical property of the gas phase. For the cases presented here, the convective contrain represents the main limitation; setting $C_{\Delta t_c}=0.15$, $C_{\Delta t_d}=0.5$ and $C_{\Delta t_{\mathcal{S}}}=1$ was found sufficient for a stable and accurate time integration. \par

\subsection{Computational set-up}
\begin{table}
\begin{subtable}{.55\textwidth}
\centering
\begin{tabular}{lccccc}
\hline
\textit{Case} & $T_{g,0}/T_c$ & \textit{Model for $\xi_g$} & $Re_{\mathcal{S}}$ & $\mathrm{We}_{\mathcal{S}}$ & $Da_v$\\
\hline
\textit{1}  & $1.50$ & CP            & $6700$ & $0.02$ & $0.8213$ \\
\textit{2}  & $1.50$ & VP$_{\rho}$   & $6700$ & $0.02$ & $0.8213$ \\
\textit{3}  & $1.50$ & VP$_{a}$      & $6700$ & $0.02$ & $0.8213$ \\
\textit{4}  & $0.75$ & CP            & $6700$ & $0.02$ & $0.1528$ \\
\textit{5}  & $0.75$ & VP$_{\rho}$   & $6700$ & $0.02$ & $0.1528$ \\
\textit{6}  & $0.75$ & VP$_{a}$      & $6700$ & $0.02$ & $0.1528$ \\
\textit{7}  & $1.00$ & VP$_{a}$      & $6700$ & $0.02$ & $0.5531$ \\
\textit{8}  & $1.50$ & VP$_{a}$      & $2800$ & $0.02$ & $0.9584$ \\
\textit{9}  & $1.00$ & VP$_{a}$      & $2800$ & $0.02$ & $0.6589$ \\
\textit{10} & $0.75$ & VP$_{a}$      & $2800$ & $0.02$ & $0.2341$ \\
\textit{11} & $1.50$ & VP$_{a}$      & $6700$ & $0.06$ & $0.8212$ \\
\textit{12} & $1.50$ & VP$_{a}$      & $6700$ & $0.10$ & $0.8213$ \\
\textit{13} & $0.75$ & VP$_{a}$      & $6700$ & $0.06$ & $0.1528$ \\
\textit{14} & $0.75$ & VP$_{a}$      & $6700$ & $0.10$ & $0.1528$ \\
\hline
\end{tabular}
\end{subtable}
\hspace{0.25 cm}
\begin{subtable}{.45\textwidth}
\vspace{-0.023 cm}
\begin{tabular}{lcc}
\hline
\textit{Group} & \textit{Definition} &\textit{Value}  \\
\hline
\textit{Pr}        & $\mu_{g,r}c_{pg,r}/k_{g,r}$           & $1.39$  \\
\textit{Sc}        & $\mu_{g,r}/(\rho_{g,r}D_{lg,r})$      & $1.66$  \\
\textit{Ste}       & $c_{pg,r}T_{g,r}/\Delta h_{lv}$       & $5.70$  \\
$\lambda_{\rho}$   & $\rho_l/\rho_{g,r}$                   & $13.15$ \\
$\lambda_{\mu}$    & $\mu_l/\mu_{g,r}$                     & $2.10$  \\
$\lambda_{c_p}$    & $c_{pl}/c_{pg,r}$                     & $2.00$  \\
$\lambda_{k}$      & $k_l/k_{g,r}$                         & $1.87$  \\
$\lambda_{M}$      & $M_l/M_g$                             & $2.58$  \\
$\Pi_{p,1}$        & $R_u/(M_gc_{pg,r})$                   & $0.18$  \\
$\Pi_{p,2}$        & $p_{th,r}M_g/(\rho_{g,r}R_uT_{g,r})$  & $1.00$  \\
$\Pi_{p,3}$        & $p_c/p_{th,r}$                        & $0.83$  \\
$\alpha_0$ $[\%]$  & $V_{l,0}/(l_xl_yl_z)$                 & $0.14$  \\
$N_{dp,0}$         & $-$                                   & $5$  \\
$RH_0$ $[\%]$      & $-$                                   & $0.0$ \\
\hline
\end{tabular}
\end{subtable}
\caption{Left: Dimensionless parameters defining the investigated cases, the initial gas temperature over the critical temperature $T_{g,0}/T_c$, the thermodynamic model employed to evaluate the gas thermophysical property, \nsc{the shear-based Reynolds number $Re_{\mathcal{S}}=\rho_{g,r}\mathcal{S}l_y^2/\mu_{g,r}$, and the shear-based Weber number $We_{\mathcal{S}}=\rho_{g,r}\mathcal{S}^2d_0^3/\sigma$ with $d_0$ the initial droplet diameter}. The vaporization Damk\"ohler number, the ratio between the turbulence time scale and the evaporation time scale in stagnant conditions, $Da_v=\tau_t/\tau_{v,L}$, is also reported (see equation~\ref{eqn:dav}). Note that in the current study $d_0/l_y=0.10$. \nsc{Right: dimensionless parameters kept constant in the current study ($N_{dp,0}$ is the initial number of droplet, $\alpha_0$ is the initial liquid volume and $RH_0$ is the initial relative humidity).}}
\label{tab:param}
\end{table}
\nsc{Given the large number of dimensionless parameter that characterizes flows involving evaporation, we focus our attention on the role of the ratio between droplet initial diameter and the Kolmogorov dissipative flow scale (tuned by varying the shear-based Reynolds number, $Re_{\mathcal{S}}$), the role of surface tension (varying the shear-based Weber number $We_{\mathcal{S}}$), the ratio between the initial gas temperature and the critical temperature, $T_{g,0}/T_c$, and the type of model employed to evaluate the thermophysical properties of the gas phase during the simulations. Concerning this last aspect, we first consider all the gas thermophyisical properties constant and computed with a proper averaging between the liquid and the gas temperature \cite[i.e., with the $1/3$ rule by][]{hubbard1975droplet} (case denoted as CP); secondly, we allow only the gas density $\rho_g$ to vary (case denoted as VP$_{\rho}$) and, finally, we allow all the thermophysical properties to vary with the local thermodynamic variables and vapour composition (case VP$_{a}$). A summary of the numerical campaign is reported in table~\ref{tab:param}, together with the remaining dimensionless physical parameters, which are all kept constant.} Note that the physical parameters reported in table~\ref{tab:param} are representative of pentane evaporating droplets in dry air at high pressure ($\sim43$ $\mathrm{bar}$). In particular, we will consider 3 values of the ratio
$T_{g,0}/T_c=0.75, 1.00, 1.50$ where $T_c$ is the critical temperature (469.69 K for pentane), which gives $T_{g,0}=354$, $470$ and $705$ K. The initial liquid temperature $T_{l,0}$ is the wet bulb temperature at $T_{g,0}$ and corresponds to $T_{l,0}=334, 388$ and $432$ K.\par
The focus of the study is the behaviour of evaporating isolated droplets in homogeneous shear turbulence and, therefore, we consider five isolated droplets (i.e., initial volume fraction $\alpha_0\approx 0.14$ $\%$ with $N_{dp,0}=5$), injected in a single-phase statistically steady-state HST field at the desired shear-based $Re_{\mathcal{S}}$ ($2800$ or $6700$). For all the cases, the computational domain has the streamwise aspect ratio, $\mathcal{AR}_{xy}=l_x/l_y\approx 2.10$, and the cross-stream ratio, $\mathcal{AR}_{zy}=l_z/l_y\approx 1.05$. \nsc{As discussed in~\cite{sekimoto2016direct}, employing such values, the effects on the flow induced by a finite-size computational box are reduced.} The domain is discretized with $1280\times 608\times 640$ grid points, thus ensuring not only an adequate resolution of the flow field (i.e., $\Delta x/\eta\approx 0.33$ for $Re_{\mathcal{S}}=2800$ and $\Delta x/\eta\approx 0.40$ for $Re_{\mathcal{S}}=6700$ where $\eta$ is the Kolmogorov length scale), but also of the droplets, whose initial resolution is $64$ points per diameter. This value is consistent with our previous study~\citep{scapin2020volume}, where we have assessed that a minimum of $50$ grid points per diameter is needed to fully resolve mass, momentum and energy transfer across the droplet interface. Note that the resolution is also sufficient to resolve the smallest scales of the two active scalar fields, $Y_{l}^{v}$ and $T$, since the associated Batchelor scales $\eta_{B,Y}=\eta/\sqrt{Sc}$ and $\eta_{B,T}=\eta/\sqrt{Pr}$, although smaller than the Kolmogorov scale, are always larger than $\Delta x$.

\section{Results}\label{sec:results}
%
%
\subsection{Comparison of the three thermodynamic models}\label{sec:3model}
\begin{figure}
    \centering
    \includegraphics[height=4.735 cm,width=8.00 cm]{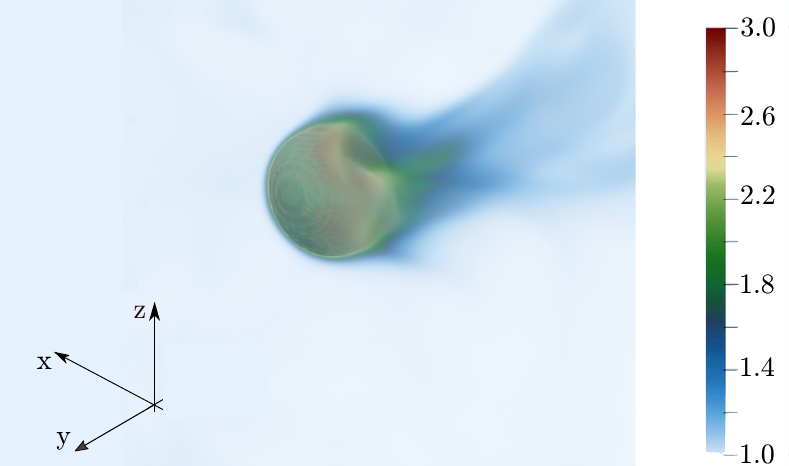}
    \put(+2.5,70){$\rho_g/\rho_{g,0}$}
    \caption{Volume rendering of the normalized gas density $\rho_{g}/\rho_{g,0}$ around a droplet for the case $Re_{\mathcal{S}}=6700$, $We_{\mathcal{S}}=0.02$ and $T_{g,0}/T_c=1.5$.}
    \label{fig:rho_g}
\end{figure}

\begin{figure*}
    \centering
    \includegraphics[width=6.0 cm,height=5.25 cm]{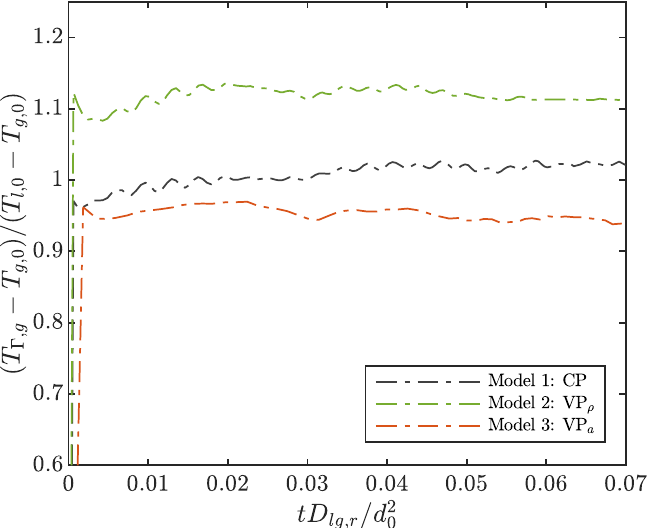}\hspace{0.5 cm}
    \includegraphics[width=6.0 cm,height=5.25 cm]{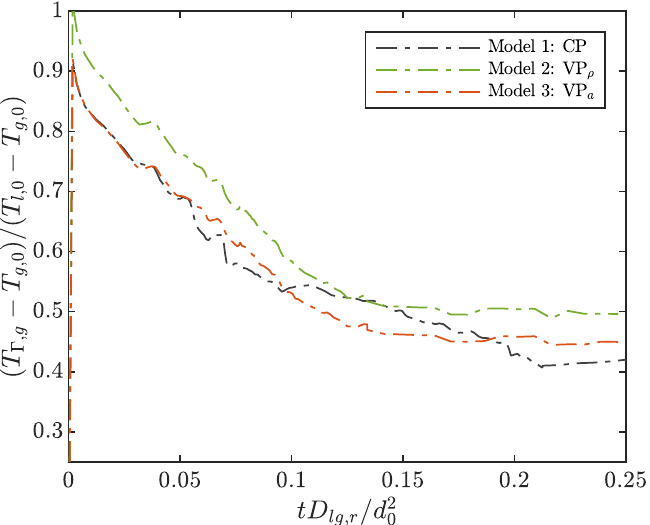}
    \put(-360,140){\small(\textit{a})}
    \put(-173,140){\small(\textit{b})}
	\caption{Normalized gas temperature at the interface as a function of the dimensionless time (based on the diffusion time scale) for temperature ratio (\textit{a}) 
      $T_{g,0}/T_c=1.50$ and (\textit{b}) $T_{g,0}/T_c=0.75$ for $Re_{\mathcal{S}}=6700$ and $We_{\mathcal{S}}=0.02$. $T_{g,0}$ and $T_{l,0}$ are the initial gas and liquid temperature.}
	\label{fig:tmpg}
\end{figure*}

\begin{figure*}
    \centering
    \includegraphics[width=6.0 cm,height=5.25 cm]{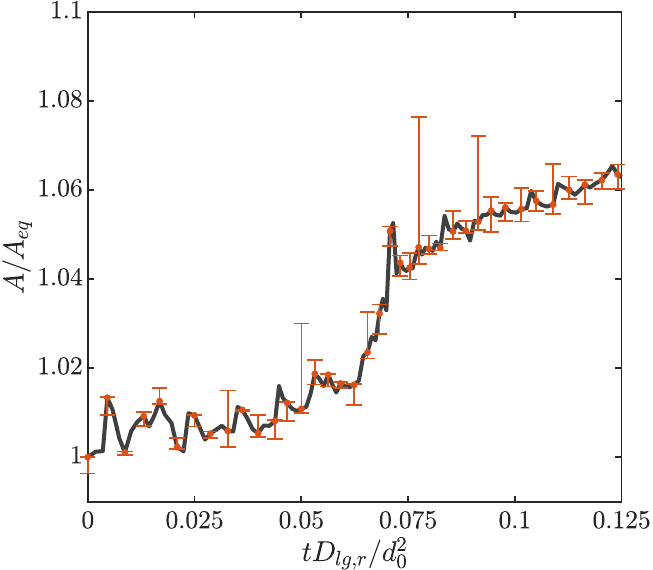}\hspace{0.5 cm}
    \includegraphics[width=6.0 cm,height=5.25 cm]{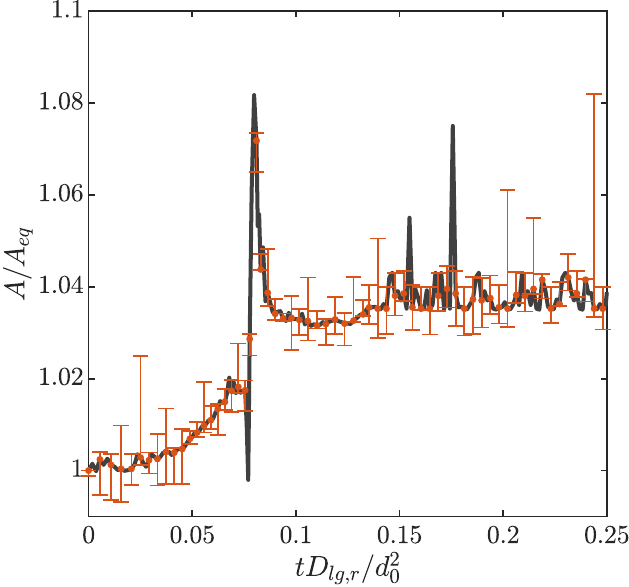}
    \put(-360,140){\small(\textit{a})}
    \put(-170,140){\small(\textit{b})}
    \caption{Ratio between the instantaneous interfacial area $A$ and the area of a spherical droplet with same volume $V$,  i.e., $A_{eq}=\pi^{1/3}(6V)^{2/3}$, for $We_{\mathcal{S}}=0.02$ and a): $T_{g,0}/T_c=1.50$ and b): $T_{g,0}/T_c=0.75$. The upper lengths of the error bar refer to the droplet with the largest $A/A_{eq}$, while the lower lengths displacement to the droplet with lowest $A/A_{eq}$. The data refers to the simulations conducted with the VP$_a$ thermodynamic models (negligible difference has been observed between the CP and the VP$_a$ model).}
	\label{fig:area}
\end{figure*}

We start our analysis by assessing the influence on the evaporation dynamics of the type of thermodynamic model employed to evaluate the gas thermophysical properties. A common approach \citep{abramzon1989droplet,lupo2019immersed,lupo2020direct} is to consider the thermophysical properties of the gas phase uniform, constant and evaluate them at an intermediate temperature, $T_m=T_{g,0}+m(T_{g,0}-T_{l,0})$. \cite{hubbard1975droplet} show that the value $m=1/3$ guarantees a good agreement between experimental results and the theoretical predictions based on the $d^2$--law~\citep{langmuir1918evaporation}. In the mathematical framework introduced above, this amounts to limiting the expansion and contraction at the interface, i.e.\ the terms $f_Y$ and $f_T$ in~\eqref{eqn:veldiv1},~\eqref{eqn:pth1} become zero. Here, we will denote the results obtained with these assumptions as CP. In many fully resolved simulations, the only thermophysical property allowed to vary is the density and this is typically done within the Oberbeck-Boussinesq approximation~\citep{piedra2015numerical}. This model, denoted VP$_{\rho}$, is considered here to asses whether it can provide accurate results. 
In typical conditions of evaporating droplets, however, other thermophysical properties may play a role. In particular, $\rho_g$ and $D_{lg}$ scale differently with the gas temperature, i.e., $\rho_g\sim p_{th}/T_g$ and $D_{lg}\sim T_g^{1.5}/p_{th}$ and both appear in the expression of the interfacial mass flux. These variations may significantly affect the evaporation dynamics, especially when the difference between the liquid and gas temperatures is large. We will therefore also consider variations of all the thermophysical properties, case VP$_{a}$. To compare the different models we consider two temperature ratios, $T_{g,0}/T_c=0.75$ and $T_{g,0}/T_c=1.50$ at $Re_{\mathcal{S}}=6700$ and the lowest Weber number under consideration in this study, $We_{\mathcal{S}}=0.02$, to limit droplet breakup and reduce the droplet deformation. 
\revA{Unless otherwise stated, the results refer to the averaged values over the five droplets initially in the computational domain and error bars are included to represent the droplet with the largest positive and negative deviation from the mean value}. Firstly, we show the square of the normalized droplet diameter, see figure~\ref{fig:d2}. From the cases at $T_{g,0}/T_c=1.50$, we observe that the complete model and the constant property model provide similar evaporation rates; when the gas density is the only varying thermophysical property, the evaporation rate is the highest. This behaviour can be attributed to the presence of colder gas around the droplets, leading to larger local gas densities (up to three times the initial gas density, see figure~\ref{fig:rho_g}) and, thus, to increased evaporation rates, as shown in figure~\ref{fig:tmpg}a where we report the time history of the surface-averaged gas temperature. Relaxing the assumption of constant liquid diffusion coefficient reduces $D_{lg} \propto T_g^{1.5}$. This counteracts the effect of the higher gas density, decreasing the overall evaporation rate, which approaches the values obtained assuming constant property values. The results at $T_{g,0}/T_c=0.75$ show a limited impact of compressibility on the evaporation dynamics, with an almost identical evaporation from the CP and VP$_{a}$ models. Once more, allowing only the gas density to vary leads to an overestimation of the evaporation rate, which is explained by a lower gas temperature and higher density at the interface, see figure~\ref{fig:tmpg}b. 
\revA{
\begin{figure*}
    \centering
    \includegraphics[width=6.0 cm,height=5.25 cm]{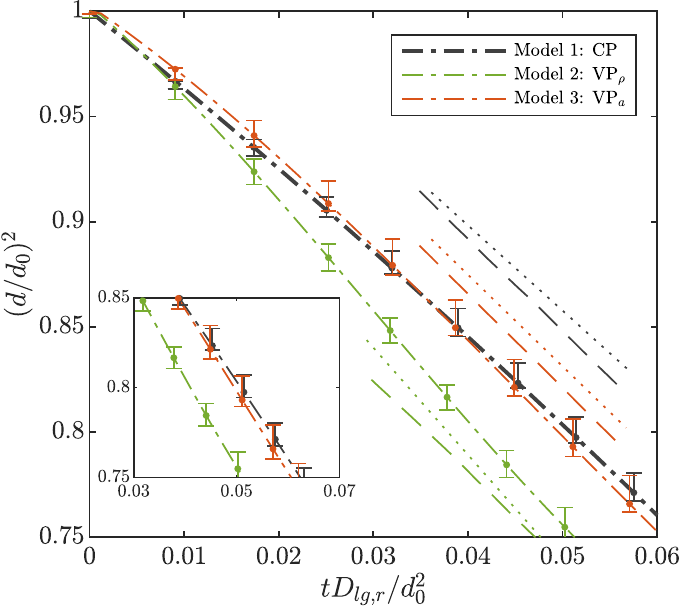}\hspace{0.5 cm}
    \includegraphics[width=6.0 cm,height=5.25 cm]{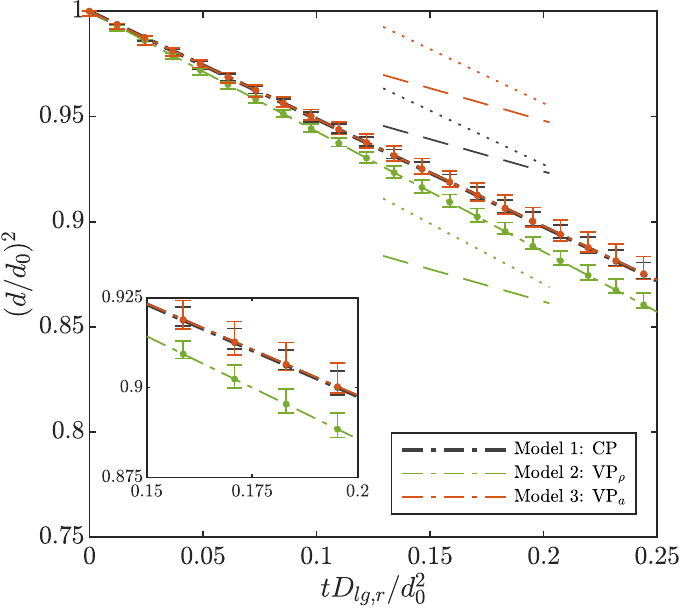}
    \put(-360,140){\small(\textit{a})}
    \put(-170,140){\small(\textit{b})}
	\caption{Normalized square droplet diameter, $(d/d_0)^2$, as a function of the dimensionless time (based on the diffusion time scale) for temperature ratio (\textit{a}) 
	$T_{g,0}/T_c=1.50$ and (\textit{b}) $T_{g,0}/T_c=0.75$ for $Re_{\mathcal{S}}=6700$  and $We_{\mathcal{S}}=0.02$. The \textit{dotted curves} indicate the slope of the curves after the initial transient (extracted from the DNS), whereas the \textit{dashed curves} is the estimation from the theoretical Fr{\"o}ssling/Ranz-Marshall correlation. 
	\revA{The length of the error bars indicates the droplet with the fastest/slowest evaporation rate among the five droplets in the simulations}.}
	\label{fig:d2}
\end{figure*}
}
\revA{Note that in all the cases and regardless of the model, the mean evaporation rate represents a good estimation of the evaporation rate of the single droplet since the magnitude of the largest positive and the negative deviations (represented by the error bars in figure~\ref{fig:d2}) is within $5$ $\%$. \\
Figure~\ref{fig:area} displays the time evolution of the  ratio between the instantaneous area $A=|\nabla\Phi|\Delta x^3$ and the area of a spherical droplet with same volume with error bars indicating the maximum and minimum values at each instance at $T_{g,0}/T_c=1.5$ and $T_{g,0}/T_c=0.75$. Note that the mean deviation form the spherical shape is always below $10$ $\%$ (over the simulated physical time) which makes the present data suitable for comparisons with existing scaling laws, which are commonly derived under the assumption of rigid and spherical droplets.
} 

In all cases, the evaporation rate follows the linear trend predicted by the $d^2$--law since the droplets are isolated. We therefore test the closure relations available in the literature against the data from the interface resolved simulations. To this end, we estimate the evaporation rate as proposed in~\cite{birouk2006current}. \nsc{First, we obtain the dimensionless evaporation rate in laminar conditions $K_{L}$ \citep{abramzon1989droplet} as}
\begin{equation}
	K_{L} = \dfrac{4\,\,Sh_{RM}}{Re_{\mathcal{S}}Sc}\dfrac{\log(1+B_M)}{\lambda_{\rho}}\mathrm{,}
  \label{eqn:klam}
\end{equation}
where the Spalding number $B_M=(\langle Y_{l,\Gamma}^v\rangle_{\Gamma}-Y_{l,\infty}^v)/(1-\langle Y_{l,\Gamma}^v\rangle_{\Gamma})$ with $\langle\rangle_{\Gamma}$ the average over the surface $\Gamma$. The Sherwood number in equation~\eqref{eqn:klam} is estimated with the Fr{\"o}ssling/Ranz-Marshall correlation~\citep{ranz1952evaporation,birouk2006current}, 
\begin{equation}
	Sh_{RM} = 2 + \dfrac{B_M}{(1+B_M)^{0.7}\log(1+B_M)}Sh_{0,RM}\mathrm{,}
  \label{eqn:sh_pp}
\end{equation}
with $Sh_{0,RM}=2+0.552Re_d^{0.50}Sc^{0.33}$. 
\revB{
In this last expression, $Sc$ is the Schmidt number and $Re_d$ is the droplet Reynolds number, defined as $Re_d= \rho_{g,r}|\mathbf{u}_{l,d}'-\mathbf{u}_{g,d}'|d/\mu_{g,r}$. Following the original reference~\citep{ranz1952evaporation} and more recent works~\citep{wang2021direct,ng2021growth}, we compute $Re_d$ using the instantaneous droplet diameter, the instantaneous droplet velocity $\mathbf{u}_{l,d}'$ and the surrounding gas velocity $\mathbf{u}_{g,d}'$ at the droplet location~\citep{ng2021growth}.
}
\nsc{Note that this correlation is based on experiments with droplets larger than the Kolmogorov scale and in presence of a mean velocity field and, is therefore, suitable for the current study.} Finally, we correct equation~\eqref{eqn:klam} for the presence of turbulence via a Damk\"ohler number~\citep{gokalp1992mass}:
\begin{equation}
	\dfrac{K}{K_L} = 0.74Da_v^{-0.115}\mathrm{.}
  \label{eqn:kturb}
\end{equation}
In the above,
\begin{equation}
	Da_v = \dfrac{\tau_t}{\tau_{v,L}}\mathrm{,}
  \label{eqn:dav}
\end{equation}
where $\tau_t=d_0^{2/3}/\langle\varepsilon_g\rangle_{M_g}^{1/3}$ is the characteristic time scale of the turbulent eddies, $d_0$ is the initial droplet diameter and $\langle\varepsilon_g\rangle_{M_g}$ is the mass averaged turbulent dissipation \cite[see][]{abramzon1989droplet,birouk2006current}. The characteristic time scale of evaporation $\tau_{v,L}$, is computed as:
\begin{equation}
	\tau_{v,L} = \dfrac{\delta_M}{V_r}\mathrm{,}
  \label{eqn:tav}
\end{equation}
\revB{
where $\delta_M=d_0/(Sh_{RM}-2)$ \citep{abramzon1989droplet} is the vapour film boundary layer thickness and $V_r$ is the Stefan flow velocity computed as $V_r=8\pi D_{lg,r}\log(1+B_M)/d_0$. Note that while equations~\eqref{eqn:klam},~\eqref{eqn:sh_pp} and the expressions for $\tau_t$ and $\delta_M$ are evaluated using also the instantaneous data from the interface-resolved simulations, the Stefan velocity $V_r$ is evaluated a-priori as it depends only on the initial thermodynamic conditions.} \\
The estimated values of $K$, dashed lines in figure~\ref{fig:d2}, display a good agreement with the interface-resolved simulations at $T_{g,0}/T_c=1.50$, especially for the CP and VP$_{a}$ models, while a significant deviation is observed at $T_{g,0}/T_c=0.75$, regardless of the thermodynamic model employed.
\nsc{Finally, it is worth noting that the evaporation rate estimated with equation~\eqref{eqn:klam} depends linearly on $Sh$, whose calculation is generally affected by a higher degree of uncertainty~\citep{lupo2020uncertainty}.}

\begin{figure*}
    \centering
    \includegraphics[width=6.0 cm,height=5.25 cm]{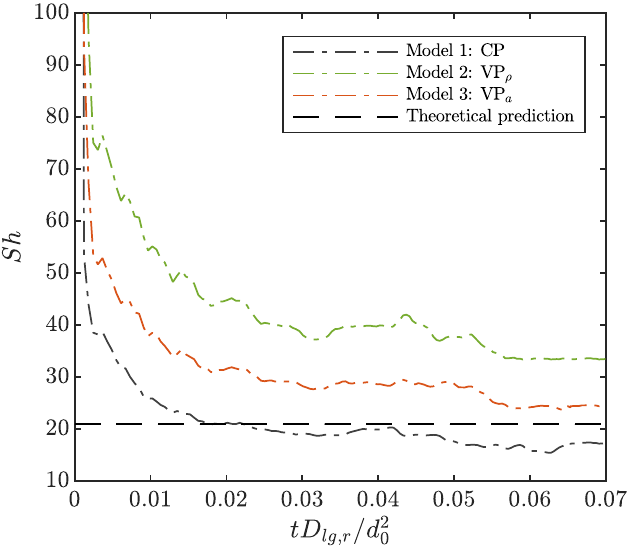}\hspace{0.5 cm}
    \includegraphics[width=6.0 cm,height=5.25 cm]{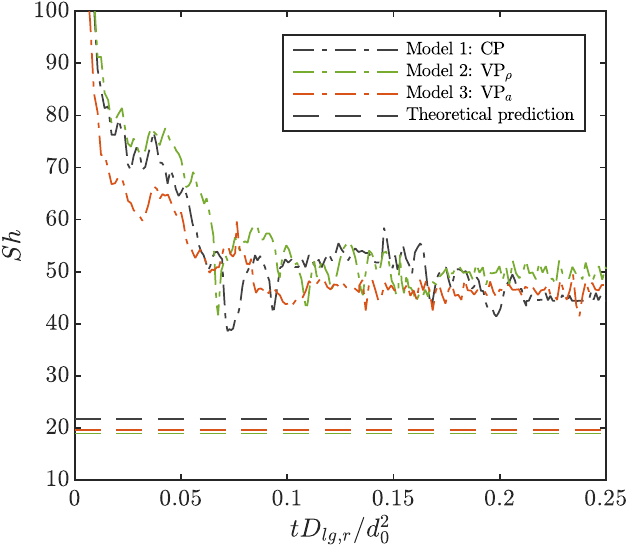}
    \put(-364,140){\small(\textit{a})}
    \put(-174,140){\small(\textit{b})}
	\caption{Temporal evolution of the Sherwood number as a function of the dimensionless time (based on the diffusion time scale) for temperature ratio (\textit{a}) 
      $T_{g,0}/T_c=1.50$ and (\textit{b}) $T_{g,0}/T_c=0.75$  and  $Re_{\mathcal{S}}=6700$  and $We_{\mathcal{S}}=0.02$.
		The dashed curves are the theoretical prediction provided from the a-priori analysis based on the Fr{\"o}ssling/Ranz-Marshall correlation. For the cases at $T_{g,0}/T_c=1.50$, only one value of $Sh$ is reported as the three models provide almost identical predictions.}
	\label{fig:sh}
\end{figure*}

Next, we  examine the temporal evolution of the Sherwood number, $Sh$, for the three different models, see figure~\ref{fig:sh}, using directly the definition
\begin{equation}
	Sh = \dfrac{\dot{m}_{t,\Gamma}d_{eq}}{\rho_{g,r}D_{lg,r}A(\langle Y_{l,\Gamma}^v\rangle_{\Gamma}-Y_{l,\infty}^v)}\mathrm{.}
  \label{eqn:sh1}
\end{equation}
Here $\dot{m}_{t,\Gamma}$ is the total interfacial mass flux, $A$ the interfacial area, $d_{eq}$ the instantaneous equivalent diameter (i.e., $d_{eq}=(6V/\pi)^3$), $\rho_{g,r}$ and $D_{lg,r}$ the reference gas density and gas-vapor diffusion coefficients, $\langle Y_{l,\Gamma}^v\rangle_{\Gamma}$ the surface-averaged vapour mass fraction at the interface and $Y_{l,\infty}^v$ the reference vapour mass fraction, i.e.\ its initial value. The Sherwood number defines the ratio between the mass transfer in the actual flow and the mass transfer by diffusion, i.e.\ evaporation in stagnant conditions, so that it can be used to quantify the effect of the flow on the evaporation. Note that since we perform a comparison among different thermodynamic models, the gas thermophysical properties in equation~\eqref{eqn:sh1} are taken equal to the reference ones (i.e., those obtained with the "1/3" rule for the CP model, those at the initial condition for VP$_{a}$ and VP$_{\rho}$ models).

Both at high and low temperatures, $Sh$ approaches an asymptotic value after the initial transient. Note that for the cases at $T_{g,0}/T_c=1.5$, the transient is faster ($\Delta tD_{lg,r}/d_0^2\approx 0.02$) than for the cases at $T_{g,0}/T_c=0.75$ ($\Delta tD_{lg,r}/d_0^2\approx 0.1$) since large temperature differences increase the evaporation rates.
For the high-temperature cases, the $Sh$ number differs significantly with the model employed, with the model VP$_{\rho}$ giving the largest value of $Sh$ for the reasons explained above.
Conversely, at lower temperature, the three models provide a similar $Sh$, confirming the limited impact on the evaporation rate of the choice of the thermodynamic models in this regime. \nsc{
	All other parameters being fixed, the Sherwood number is higher at the lower temperature ratio, $T_{g,0}/T_c=0.75$ than for $T_{g,0}/T_c=1.50$, regardless of the thermodynamic model used.} Recalling that the Sherwood number is the ratio between the actual mass transfer and the mass transfer by diffusion, the increase of evaporation rate due to the background turbulence is thus larger at low temperatures. For the highest temperature ratio, the enhancement induced by the turbulence is lower since evaporation is mainly driven by the large temperature difference and by the larger values of $(\langle Y_{l,\Gamma}^v\rangle_{\Gamma}-Y_{l,\infty}^v)$. \par
As a final remark, note that the Sherwood number from the direct numerical simulation (DNS) agrees well with the estimation from the Fr\"{o}ssling/Ranz-Marshall correlation at high temperature (cf. models CP and VP$_{a}$ against the dashed line the figure~\ref{fig:sh}). Conversely, at lower temperature, the estimated value is much lower than the one extracted from the direct numerical simulations, indicating that the Fr\"{o}ssling/Ranz-Marshall correlation underpredicts convective effects due to turbulence at low evaporation rates, i.e.\ when the evaporation time becomes longer than the turbulence time scale ($Da_v \lesssim 0.2$ instead of $Da_v \approx 1$ as shown in table \ref{tab:param}). \nsl{This finding is consistent with the recent investigations in HIT performed by~\cite{mees2020statistical} and suggests the need of including the turbulent effects (via the vaporization Damk\"ohler number) in the available correlations for the Sherwood number and more in general in the evaporation models~\citep{lupo2020direct}. \par
}
%
%
\subsection{Effects of the shear-based Reynolds number}\label{sec:re_effect} 
\begin{figure*}
    \centering
    \includegraphics[width=6.0 cm, height=5.25 cm]{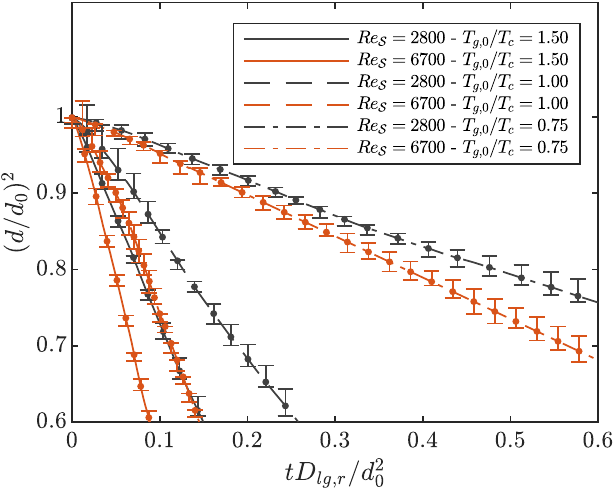}\hspace{0.5 cm}
    \includegraphics[width=6.0 cm, height=5.25 cm]{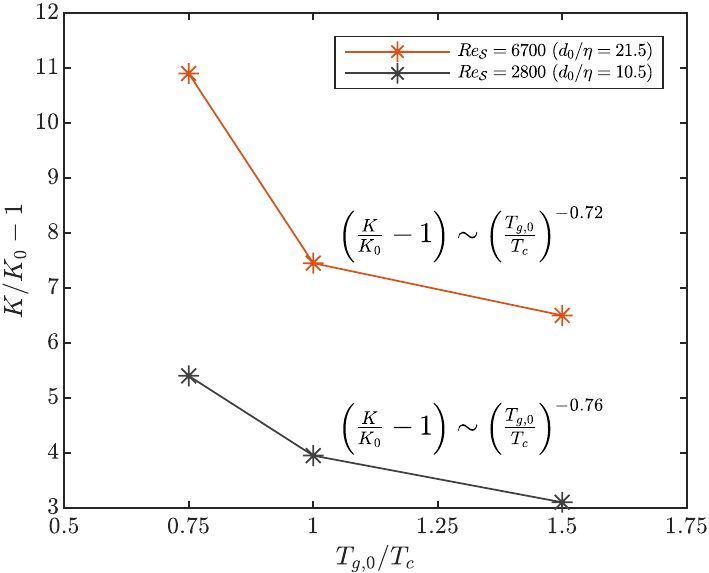}
    \put(-360,140){\small(\textit{a})}
    \put(-174,140){\small(\textit{b})}
	\caption{a) Normalized square droplet diameter $(d/d_0)^2$ as a function of the diffusion time scale $tD_{lg}/d_0^2$. b) \nsc{Ratio between the turbulent evaporation rate $K$ and the one at stagnant conditions,} $K_0=(8/\lambda_{\rho})\log(1+B_M)$ with $B_M=(Y_{l,\Gamma}^v-Y_{l,\infty}^v)/(1-Y_{l,\Gamma}^v)$. The data refer to the cases at $Re_{\mathcal{S}}=2800-6700$ and $T_{g,0}/T_c=1.5, 1.00$ and $0.75$. \revA{The length of the error bars indicates the droplet with the fastest/slowest evaporation rate among the five droplets in the simulations.}} 
    \label{fig:d2_diff_re}
\end{figure*}

We now consider the effect of the shear-based Reynolds number on the evaporation rate. Reducing from $Re_{\mathcal{S}}=6700$ to $Re_{\mathcal{S}}=2800$ results in a reduction of the ratio between the initial droplet diameter and the Kolmogorov length scale from $d_0/\eta=21.5$ to $d_0/\eta=10.5$. For this analysis, we assume $We_{\mathcal{S}}=0.02$, three temperature ratios, $T_{g,0}/T_c=0.75-1.00-1.50$ and employ the complete model (i.e., VP$_a$). 
We first examine the reduction of the droplet diameter, $(d/d_0)^2$, as a function of the diffusion time scale $tD_{lg,r}/d_0^2$, see figure~\ref{fig:d2_diff_re}a. \nsc{In all cases, droplets with larger $d_0/\eta$ (i.e. higher $Re_{\mathcal{S}}$) evaporate faster, in agreement with previous experimental studies in homogeneous isotropic turbulence \citep{verwey2018fuel,verwey2020experimental} and in the presence of a mean flow \citep{marti2017evaporation}. This is explained  by the enhanced surface vapour gradient and faster dispersion of the vapour around the droplet for larger $d_0/\eta$, related to the fact that larger flow structures are more energetic. Note that a relative increase of the evaporation rate, quantified as the ratio between the evaporation rate $K$ extracted from the DNS and the one computed in stagnant conditions, $K_0$, occurs at all temperatures as reported in figure~\ref{fig:d2_diff_re}b} \revB{and it scales with the dimensionless gas temperature with similar exponents for the two different  $d_0/\eta$ investigated, i.e. $~(T_{g,0}/T_c)^{-0.76}$ for $d_0/\eta=21.5$ and $(T_{g,0}/T_c)^{-0.72}$ for $d_0/\eta=10.5$}. The experimental correlations in \cite{birouk2006current} suggests that, at high temperature and pressure i.e.,\ when the evaporation is much faster than the turbulent time scale, the evaporation is only weakly dependent on the droplet size $d_0/\eta$. This is consistent with our results, where the difference in $K$ for the two droplet sizes examined decreases when increasing the temperature; however, we do not observe the rate of evaporation to become independent of the size in the parameter range considered.

\begin{figure}
  \centering
  \includegraphics[width=6.0 cm, height=5.25 cm]{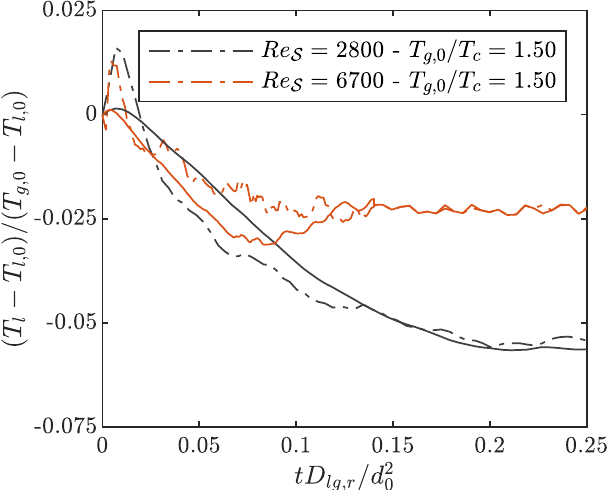}
  \includegraphics[width=6.0 cm, height=5.25 cm]{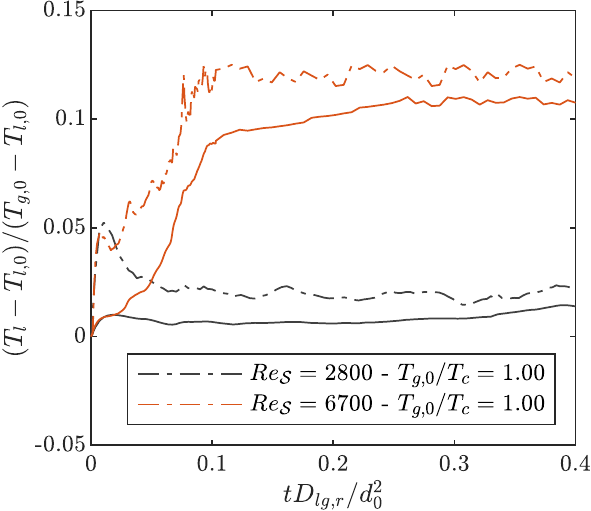}
  \put(-351,140){\small(\textit{a})}
  \put(-174,140){\small(\textit{b})}
  \vspace{0.25 cm}
  \includegraphics[width=6.0 cm, height=5.25 cm]{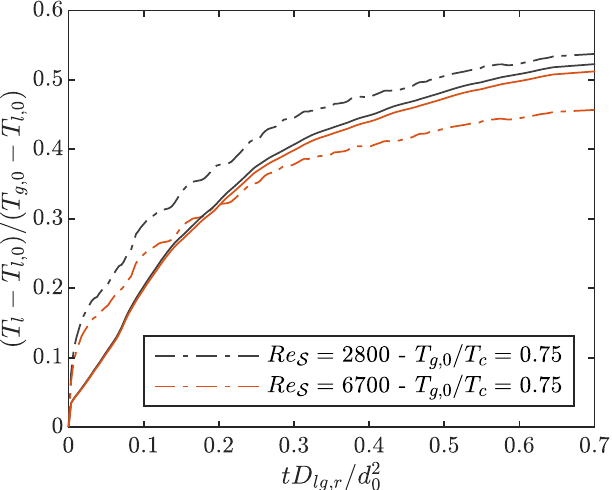}
  \put(-175,+140){\small(\textit{c})}
  \caption{Dimensionless liquid temperature at the interface $(T_{l,\Gamma}-T_{l,0})/(T_{g,0}-T_{l,0})$ (dotted line) and average liquid temperature $(T_{l,V}-T_{l,0})/(T_{g,0}-T_{l,0})$ (continuous line) at a) $T_{g,0}/T_c=1.5$ , b) $T_{g,0}/T_c=1.0$ and c) $T_{g,0}/T_c=0.75$.}
  \label{fig:ts_nu_ht}
\end{figure}

The variation of the evaporation rate with $d_0/\eta$ significantly affects the liquid temperature at the interface $T_{\Gamma,l}$ (computed as the surface average over the gas-liquid interface, $T_{\Gamma,l}=(1/\Gamma)\int_{\Gamma} Td\Gamma$) and the mean liquid temperature $T_{V,l}$ (computed as the volume average over the liquid, $T_{V,l}=(1/V_l)\int_{V_l} TdV_l$). This is shown in figure~\ref{fig:ts_nu_ht} for the three temperatures under investigation. 
At the highest temperature, panel \ref{fig:ts_nu_ht}a, both $T_{\Gamma,l}$ and $T_{V,l}$ decrease with respect to the initial values due to the strong cooling caused by the
evaporation. \nsc{However, due to the heat transfer enhancement} at higher $Re_{\mathcal{S}}$, $T_{\Gamma,l}$ remains greater than $T_{V,l}$ for droplets at larger $d_0/\eta$, whereas  $T_{\Gamma,l}$ is lower than $T_{V,l}$ for the smaller $d_0/\eta$ considered. Clearly, the two temperatures eventually converge to a similar value once the thermal gradients inside the liquid reduce. For the case with intermediate temperature, see panel \ref{fig:ts_nu_ht}b, the cooling due to evaporation is not sufficiently high to counteract the heat released from the gas and the two temperatures reach a regime value larger than the initial one. The final values of 
$T_{\Gamma,l}$ and $T_{V,l}$ are higher for the droplets with larger $d_0/\eta$ (larger Reynolds number). We also note that the interface region is warmer than the droplet bulk  (i.e., $T_{\Gamma,l}>T_{V,l}$) for both droplet sizes. At the lowest temperature, panel c of the figure \ref{fig:ts_nu_ht}, both $T_{\Gamma,l}$ and $T_{V,l}$ increase with time, when the evaporation is too slow to cool the droplets. Unlike the previous two cases, the values of $T_{\Gamma,l}$ and $T_{V,l}$ at $d_0/\eta=21.5$ are always lower than the corresponding ones at $d_0/\eta=10.5$, due to the higher evaporation rate at the highest $Re_{\mathcal{S}}$. For this case, the faster evaporation also changes the transient of $T_{\Gamma,l}$, which at the beginning is higher than $T_{V,l}$ while for $tD_{lg,r}/d_0>0.2$ becomes lower.
%


We conclude this section by analysing the relative importance of the convective and conductive heat and mass vapour fluxes, $\mathcal{F}_{T,i=c,d}$ and $\mathcal{F}_{Y,i=c,d}$, in the gas region around the droplets and inside the liquid region. 
These are computed as an integral over the control surfaces $S_T$ and $S_{Y}$,
\begin{equation}
\begin{aligned}
  &\mathcal{F}_{T,d} = -\dfrac{1}{S_T}\int_{S_T} k\nabla T\cdot\mathbf{n}_{\Gamma}dS_T 
  \mathrm{,} \\
  &\mathcal{F}_{T,c} =  \dfrac{1}{S_T}\int_{S_T} \rho c_p\left(\mathbf{u}' T+\mathcal{S}z\mathbf{e}_xT\right)\cdot\mathbf{n}_{\Gamma}dS_T\mathrm{,} \\
	&\mathcal{F}_{Y,d} = -\dfrac{1}{S_Y}\int_{S_Y} \rho_gD_{lg}\nabla Y_{l}^{v}\cdot\mathbf{n}_{\Gamma}dS_Y
  \mathrm{,} \\
	&\mathcal{F}_{Y,c} =  \dfrac{1}{S_Y}\int_{S_Y} \rho_g\left(\mathbf{u}' Y_{l}^{v}+\mathcal{S}z\mathbf{e}_xY_l^v\right)\cdot\mathbf{n}_{\Gamma}dS_Y\mathrm{.} \\
\end{aligned}
\label{eqn:fluxes}
\end{equation}
In the gas phase, $S_T$ and $S_Y$ are surfaces conforming to the droplet shape (quasi-spherical at this low value of $We_{\mathcal{S}}$)
at a distance 
approximately equal to the thermal and vapour mass fraction boundary layer thicknesses $\delta_T$ and $\delta_Y$.  
These are  determined as the distance from the interface where $(T(\delta_T)-T_{\Gamma})/(T_{g,0}-T_{\Gamma})=0.99$ and $(Y_l^v(\delta_Y)-Y_{l,\Gamma}^v)/(Y_{l,0}^v-Y_{l,\Gamma}^v)=0.99$ as in~\cite{ni2021numerical}, moving normal to the interface thanks to a level-set function, reconstructed from the VoF field. 
We evaluate $\mathcal{F}$ as in equations~\eqref{eqn:fluxes} after the evaporation reaches a steady condition (i.e., $d^2$ regime)
and average in time over a time interval $\Delta t=[0.20,0.50,0.60](tD_{lg,r})/d_0^2$ for the high, intermediate and low temperature ratios. 
To capture the dominant transport mechanism in the liquid phase, $S_T$ is defined at a distance from the interface equal to half the instantaneous droplet radius.

\begin{figure*}
    \centering
    \includegraphics[width=5.5 cm,height=5.0 cm]{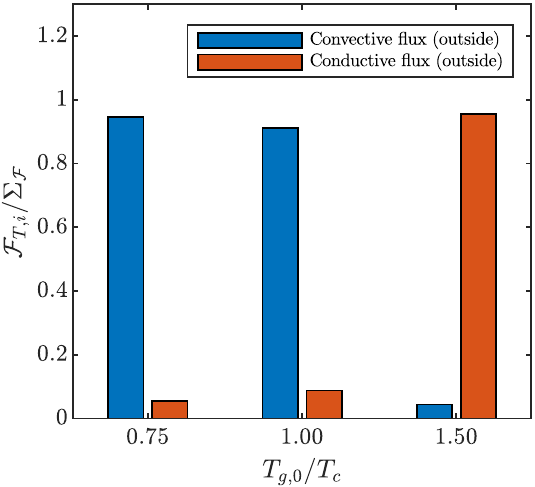}\hspace{0.5 cm}
    \includegraphics[width=5.5 cm,height=5.0 cm]{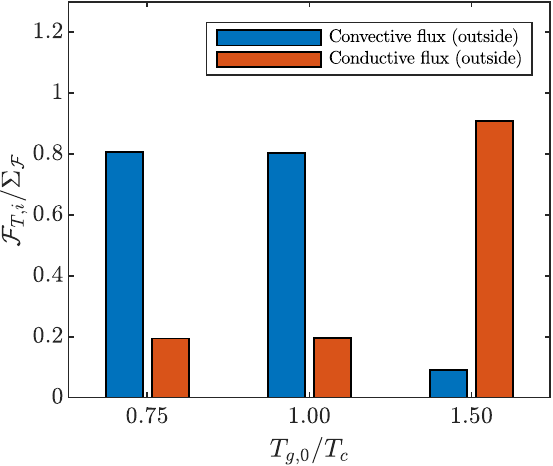}
    \put(-334,132){\small(\textit{a})}
    \put(-164,132){\small(\textit{b})}
	\caption{Conductive and convective heat fluxes $\mathcal{F}_T$ (over the sum of the two, $\Sigma_{\mathcal{F}}$), $\mathcal{F}_{T,i=c,d}$, in the gas region for $T_{g,0}/T_{c}=0.75-1.00-1.50$ and: a) $Re_{\mathcal{S}}=6700$, b) $Re_{\mathcal{S}}=2800$. At $Re_{\mathcal{S}}=6700$, $\Sigma_{\mathcal{F},  0.75}=0.08 \Sigma_{\mathcal{F},1.50}$ and $\Sigma_{\mathcal{F},1.00}=0.14\Sigma_{\mathcal{F},1.50}$. At $Re_{\mathcal{S}}=2800$, $\Sigma_{\mathcal{F},0.75}=0.04 \Sigma_{\mathcal{F},1.50}$ and $\Sigma_{\mathcal{F},1.00}=0.13\Sigma_{\mathcal{F},1.50}$ (In $\Sigma_{\mathcal{F},x}$, $x$ is the temperature ratio).}
	\label{fig:hist_tmp_out}
\end{figure*}

\begin{figure*}
    \centering
    \includegraphics[width=5.5 cm,height=5.0 cm]{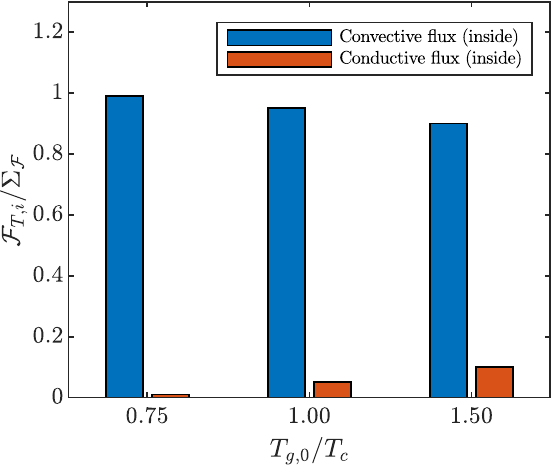}\hspace{0.5 cm}
    \includegraphics[width=5.5 cm,height=5.0 cm]{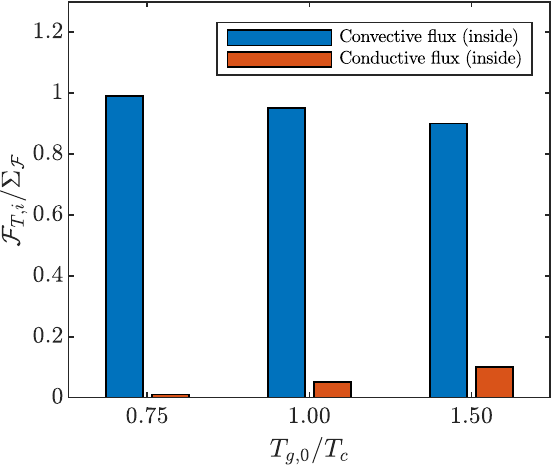}
    \put(-334,132){\small(\textit{a})}
    \put(-164,132){\small(\textit{b})}
	\caption{Conductive and convective heat fluxes $\mathcal{F}_T$ (over the sum of the two, $\Sigma_{\mathcal{F}}$), $\mathcal{F}_{T,i=c,d}$, in the liquid region for $T_{g,0}/T_{c}=0.75-1.00-1.50$ and: a) $Re_{\mathcal{S}}=6700$, b) $Re_{\mathcal{S}}=2800$. At $Re_{\mathcal{S}}=6700$, $\Sigma_{\mathcal{F},0.75}=0.12\Sigma_{\mathcal{F},1.50}$ and $\Sigma_{\mathcal{F},1.00}=0.24\Sigma_{\mathcal{F},1.50}$. At $Re_{\mathcal{S}}=2800$, $\Sigma_{\mathcal{F},0.75}=0.11\Sigma_{\mathcal{F},1.50}$ and $\Sigma_{\mathcal{F},1.00}=0.48\Sigma_{\mathcal{F},1.50}$ (In $\Sigma_{\mathcal{F},x}$, $x$ is the temperature ratio).}
	\label{fig:hist_tmp_in}
\end{figure*}

\begin{figure*}
    \centering
    \includegraphics[width=5.5 cm,height=5.0 cm]{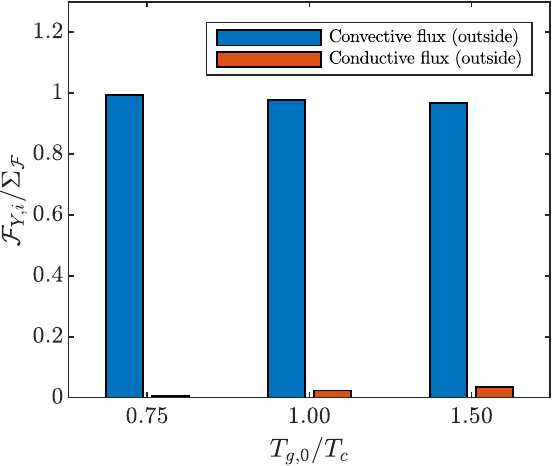}\hspace{0.5 cm}
    \includegraphics[width=5.5 cm,height=5.0 cm]{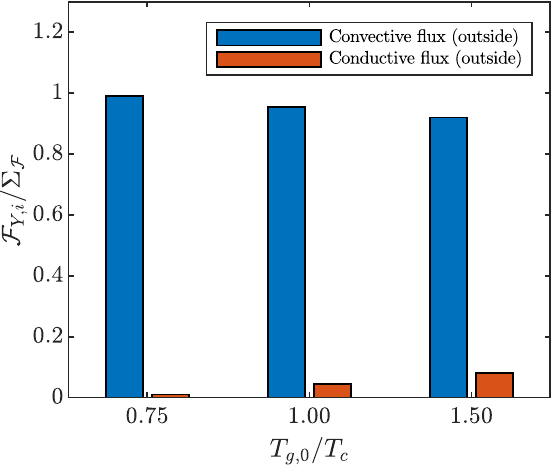}
    \put(-334,132){\small(\textit{a})}
    \put(-164,132){\small(\textit{b})}
	\caption{Conductive and convective vapour mass fluxes $\mathcal{F}_Y$ (over the sum of the two, $\Sigma_{\mathcal{F}}$), $\mathcal{F}_{Y,i=c,d}$, in the gas region for $T_{g,0}/T_{c}=0.75-1.00-1.50$ and: a) $Re_{\mathcal{S}}=6700$, b) $Re_{\mathcal{S}}=2800$. At $Re_{\mathcal{S}}=6700$, $\Sigma_{\mathcal{F},0.75}=0.58\Sigma_{\mathcal{F},1.50}$ and $\Sigma_{\mathcal{F},1.00}=0.89\Sigma_{\mathcal{F},T_{g,0}/T_c=1.50}$. At $Re_{\mathcal{S}}=2800$, $\Sigma_{\mathcal{F},0.75}=0.56\Sigma_{\mathcal{F},T_{g,0}/T_c=1.50}$ and $\Sigma_{\mathcal{F},1.00}=0.87\Sigma_{\mathcal{F},1.50}$ (In $\Sigma_{\mathcal{F},x}$, $x$ is the temperature ratio).}
	\label{fig:hist_sca_out}
\end{figure*}
The relative importance of the conductive and convective heat fluxes in the gas region are shown in figure~\ref{fig:hist_tmp_out}: increasing the ratio $T_{g,0}/T_c$, the main transport mechanism changes from convection to conduction, as already anticipated in section~\ref{sec:3model} when discussing the Sherwood number  $Sh$. The same applies to lower $Re_{\mathcal{S}}$, where we observe a slightly higher conductive contribution for $T_{g,0}/T_c=1.5$. In the liquid region, fig~\ref{fig:hist_tmp_in}, the heat transport is mainly driven by convection in all cases. In agreement with previous experimental results~\citep{wong1992internal,pinheiro2019ethanol}, the dominant transport mechanism is weakly affected by a change of the droplet Reynolds number, here varied by changing $Re_{\mathcal{S}}$, since the motion of the liquid inside the droplet is not significantly affected by the intensity of the gas turbulence. Finally, the vapour fluxes reported in figure~\ref{fig:hist_sca_out} show that $Y_l^v$ is predominantly transported by convection for the two Reynolds numbers investigated, mass diffusion being relevant only at the highest temperatures.

\subsection{Effects of the shear-based Weber number}
At last, we consider the effect of the shear-based Weber number on the evaporation rate. For this analysis, three Weber numbers $We_{\mathcal{S}}=0.02-0.06-0.10$ and two temperature ratios $T_{g,0}/T_c=0.75-1.5$ are considered and, as for the previous section, the complete thermodynamic model (i.e., VP$_a$) is employed to evaluate the gas thermophysical properties. 
\begin{figure}
  \centering
  \includegraphics[width=6.0 cm, height=5.25 cm]{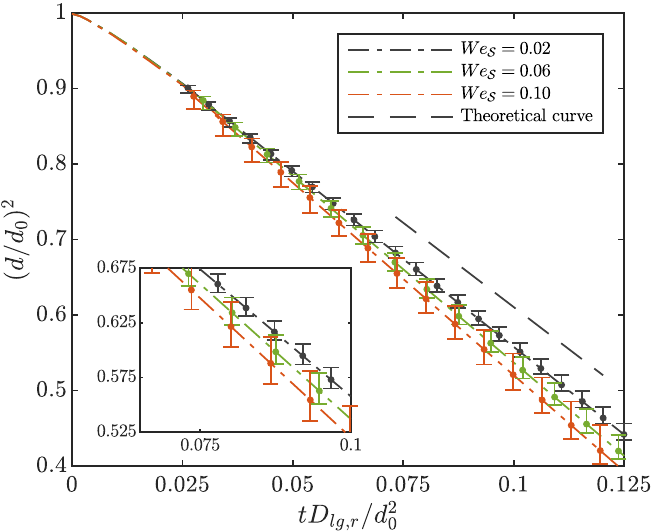}
  \includegraphics[width=6.0 cm, height=5.25 cm]{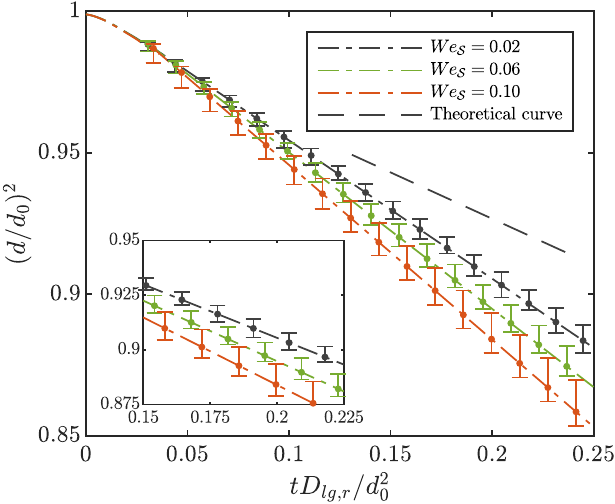}
  \put(-351,140){\small(\textit{a})}
  \put(-174,140){\small(\textit{b})}\\
  \vspace{0.25 cm}
  \includegraphics[width=6.0 cm, height=5.25 cm]{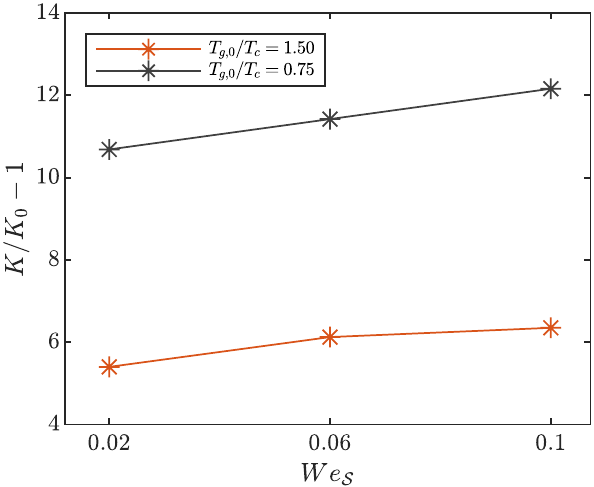}
  \put(-175,+140){\small(\textit{c})}
  \caption{Time history of $(d/d_0)^2$ when varying the Weber number, $We_{\mathcal{S}}=0.02-0.06-0.10$ and for temperature ratios a): $T_{g,0}/T_c=1.50$ and b): $T_{g,0}/T_c=0.75$. \nsc{The black dashed curves represent the theoretical prediction of the actual evaporation rate obtained with the procedure explained in section~\ref{sec:3model}}. \revA{The length of the error bars (included for $tD_{lg,r}/d_0^2>0.025$ and $tD_{lg,r}/d_0^2>0.04$, respectively) indicates the droplet with the fastest/slowest evaporation rate among the five droplets in the simulations.} c) Evaporation enhancement (i.e., $K/K_0$) as function of $We_{\mathcal{S}}$ for $Re_{\mathcal{S}}=6700$ and $T_{g,0}/T_c=0.75-1.50$.}
  \label{fig:d2_diff_we}
\end{figure}
Figure~\ref{fig:d2_diff_we}a-b) reports the time evolution of $d^2$ for the three values of $We_{\mathcal{S}}$ under investigation at high and low temperatures, respectively. Increasing $We_{\mathcal{S}}$ leads to a higher evaporation rate, due to the increase of the surface area available for mass transfer as a consequence of deformation. \revB{Note, also, that the increase in the evaporation rate with $We_{\mathcal{S}}$ is more pronounced at lower temperature (see~\ref{fig:d2_diff_we}c), which can be explained by the fact that the evaporation is faster at higher temperatures, the droplets are therefore smaller, which leads to a limited deformation with respect to the cases at the same $We_{\mathcal{S}}$ and $T_{g,0}/T_c=0.75$. In other words, the effective Weber number based on the droplet diameter becomes quickly smaller at higher temperatures, so differences in nominal surface tension are compensated by the reduced droplet size.}
To quantify deformation while accounting for the decrease of the liquid volume, we measure the deviation from the initial spherical shape as the ratio between the instantaneous interfacial area $A$ (computed numerically as $A=|\nabla\Phi|\Delta x^3$) and the area of a spherical droplet with the same volume $V$, i.e., $A_{eq}=\pi^{1/3}(6V)^{2/3}$. This calculation is performed for each droplet $q$ separately and each contribution $A_q/A_{eq,q}$ averaged according to the instantaneous number of droplets, $N_{dp}$:
\begin{equation}
  \dfrac{A}{A_{eq}}=\dfrac{1}{N_{dp}}\displaystyle{\sum_{q=1}^{N_{dp}}\dfrac{A_q}{A_{eq,q}}}\mathrm{.}
\end{equation} 
\begin{figure*}
    \centering
    \includegraphics[width=6.0 cm, height=5.25 cm]{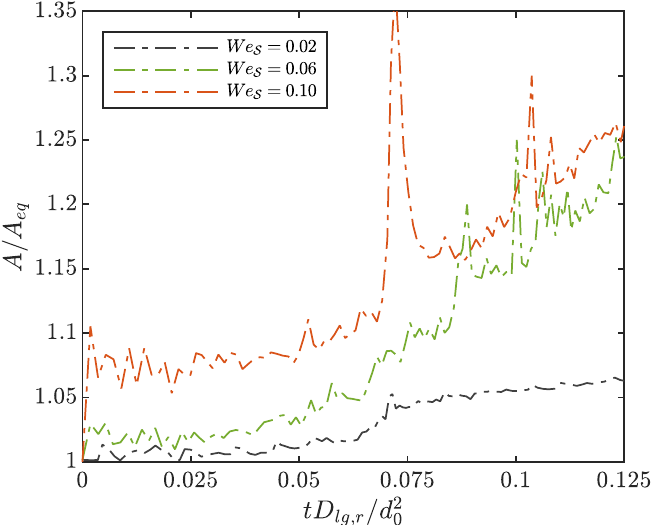}
    \includegraphics[width=6.0 cm, height=5.25 cm]{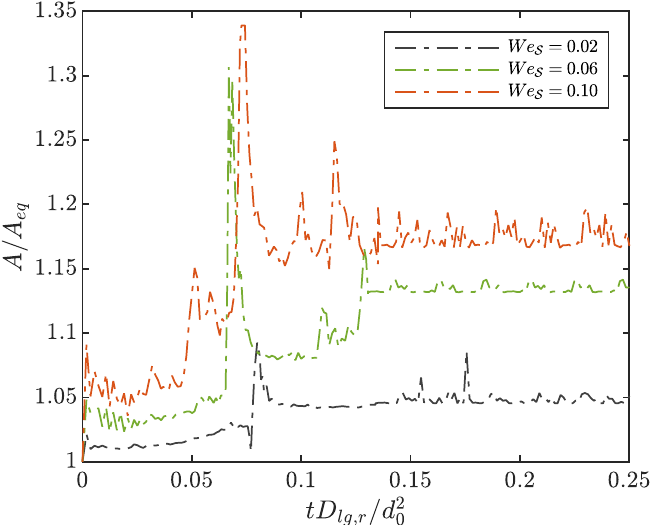}
    \put(-351,140){\small(\textit{a})}
    \put(-174,140){\small(\textit{b})}
    \caption{Ratio between the instantaneous interfacial area $A$ and  the area of a spherical droplet with same volume $V$,  i.e., $A_{eq}=\pi^{1/3}(6V)^{2/3}$,  for $We_{\mathcal{S}}=0.02-0.06-0.10$ and a): $T_{g,0}/T_c=1.50$ and  b): $T_{g,0}/T_c=0.75$.}
    \label{fig:area_diff_we}
\end{figure*}

\begin{figure}
  \centering
  \includegraphics[width=6.5 cm,height=5.2 cm]{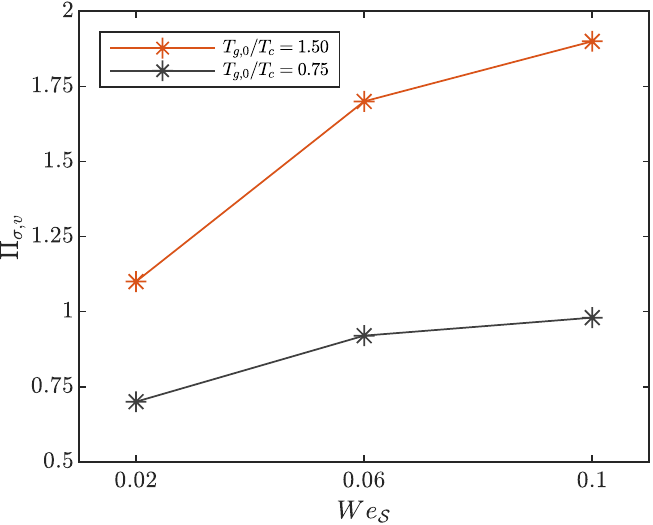}
  \caption{Ratio between the deformation time scale and the evaporation time scale (in turbulent condition) as a function of $We_{\mathcal{S}}$ for $T_{g,0}/T_c=0.75$ and $1.50$.}
  \label{fig:pi_sigma_n}
\end{figure}

The time evolution of $A/A_{eq}$ is reported in figure~\ref{fig:area_diff_we}: as expected, the ratio $A/A_{eq}$ increases with $We_{\mathcal{S}}$. For $T_{g,0}/T_c=1.5$, the data in the figure also display two peaks: at $tD_{lg,r}/d_0^2\approx 0.075-0.10$ for $We_{\mathcal{S}}=0.10$ and $tD_{lg,r}/d_0^2\approx 0.085-0.10$ for $We_{\mathcal{S}}=0.06$. These are associated to the droplet breakup as we have $N_{dp}=7$ droplets for $We_{\mathcal{S}}=0.06$ and $N_{dp}=9$ for $We_{\mathcal{S}}=0.10$ at the end of the simulations. For $We_{\mathcal{S}}=0.02$, no breakup events have been observed within the simulation time. 
For $T_{g,0}/T_c=0.75$, given the lower evaporation rate and consequently the higher effective Weber number, breakup events have been observed for all the cases in the interval $0.05<tD_{lg,r}/d_0^2<0.125$, finally yielding $N_{dp}=7-9-12$ for $We_{\mathcal{S}}=0.02-0.06-0.10$. \\
Interestingly, for the case $T_{g,0}/T_c=0.75$, $A/A_{eq}$ approaches a statistically stationary value for $tD_{lg}/d_0^2>0.125$, whereas a regime value is not reached in the investigated time window for $T_{g,0}/T_c=1.50$. This behaviour is interpreted by comparing the time scale of deformation as dictated by surface tension and applied shear, $\tau_{\sigma}$, with the time scale of evaporation extracted from the simulations, $\tau_{v}$,
\begin{equation}
  \Pi_{\sigma v}=\dfrac{\tau_{\sigma}}{\tau_v}\mathrm{,}
\end{equation}
where \revB{$\tau_{\sigma}=\rho_{g,r}(\mathcal{S}d_0)d_0^2/\sigma$} and $\tau_v=d_0^2/K$, with $K$ the evaporation at statically steady state, see figure~\ref{fig:pi_sigma_n}. At the high temperature ratio, evaporation is faster than deformation (i.e., $\tau_{\sigma}>\tau_v$) and the surface tension does not have time to adjust the droplet shape after the mass losses due to the differential evaporation across the interface. An increasing deviation from the spherical shape is therefore observed in time. Conversely, at low temperature, the deformation time is comparable or lower than the evaporation time (i.e., $\tau_{\sigma}<\tau_v$) and the droplet shape can compensate for the local deformations induced by the evaporation mass flux (note that the droplet tends to a constant deformation, increasing with $We_{\mathcal{S}}$ and not to the spherical shape, i.e.\ $A/A_{eq}>1$, as we have an imposed shear).
From the data in figure~\ref{fig:pi_sigma_n}, we also notice that for a fixed $T_{g,0}/T_c$, increasing $We_{\mathcal{S}}$ accelerates the evaporation time scale (i.e., higher $\Pi_{\sigma v}$), consistently with what observed in figure~\ref{fig:d2_diff_we}.

\begin{figure*}
    \centering
    \includegraphics[width=12.60 cm, height=4.335 cm]{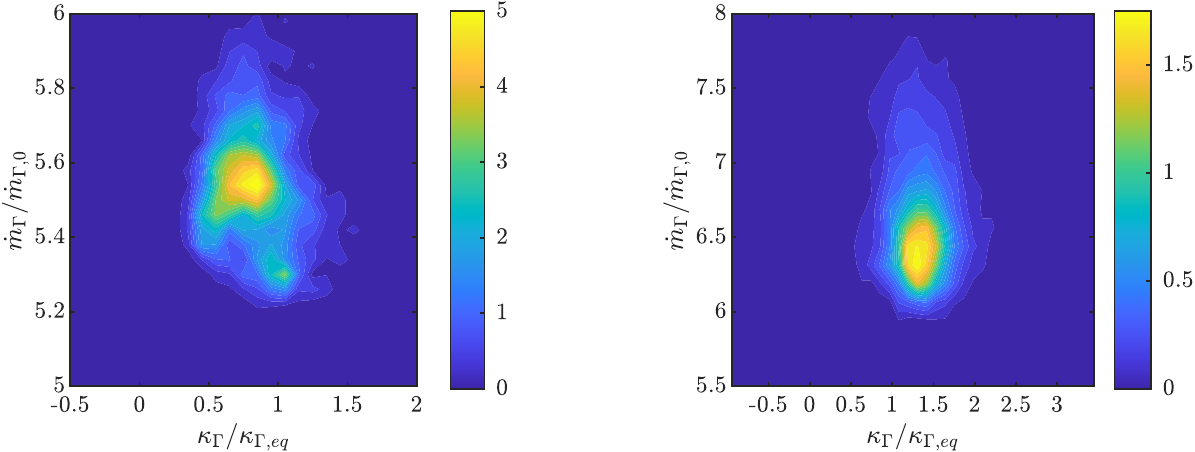}
    \put(-361,114){\small(\textit{a})}
    \put(-163,114){\small(\textit{b})}
    \caption{Joint p.d.f. of the dimensionless mass flux $\dot{m}_{\Gamma}/\dot{m}_{\Gamma,0}$ and the normalized interfacial curvature $\kappa_{\Gamma}/\kappa_{\Gamma,eq}$ for $Re_{\mathcal{S}}=6700$, $T_{g,0}/T_c=1.5$ and a) $We_{\mathcal{S}}=0.02$, b) $We_{\mathcal{S}}=0.10$.}
    \label{fig:jpdf_mk_lw_hr}
\end{figure*}

Given the variation of the evaporation rate with the $We_{\mathcal{S}}$, it is worth investigating whether the evaporation flux at the droplet surface is correlated to the local curvature. This has been already assessed in laminar flows analytically for evaporating droplets of spheroidal shapes~\citep{tonini2013exact}, and experimentally for sessile droplets~\citep{saenz2017dynamics}. In turbulent flows this analysis is missing and is performed here by computing the joint probability density function (p.d.f.) of the normalized interfacial mean curvature $\kappa_{\Gamma}/\kappa_{\Gamma,eq}$ and the dimensionless interfacial mass flux $\dot{m}_{\Gamma}/\dot{m}_{\Gamma,0}$, where $\kappa_{\Gamma,eq}$ is the curvature of a spherical droplet with same volume $V$ (i.e., 
$\kappa_{\Gamma,eq}=4/d_{eq}$ with $d_{eq}=(6V/\pi)^{1/3}$) and $\dot{m}_{\Gamma,0}$ is the interfacial mass flux for a purely diffusion-dominated evaporation process, i.e., 
\begin{equation}
  \dot{m}_{\Gamma,0}=8\pi\rho_{g,r}D_{lg,r}\log(1+B_M)/d_0\mathrm{.}
\end{equation}
Note that $\dot{m}_{\Gamma}$ is evaluated using the expression~\eqref{eqn:int_mfx} in all the nodal points cut by the interface, whereas $\kappa_{\Gamma}$ is computed directly from its definition, i.e., $\kappa_{\Gamma}=\nabla\cdot \mathbf{n}_{\Gamma}$. The reconstructed level-set function is employed for a more accurate estimation of $\kappa_{\Gamma}$. \\
From the results at high temperature (i.e., $T_{g,0}/T_c=1.5$) reported in figure~\ref{fig:jpdf_mk_lw_hr},  we do not note a clear relation between $\dot{m}_{\Gamma}$ and $\kappa_{\Gamma}$, yet there is a relatively broad distribution of mass fluxes over the droplet surface, which can be attributed to the local variations of vapour concentrations in the gas phase induced \revA{by the turbulence and by the flow anisotropy. Indeed, the regions whose local outward normal vector is parallel to the mean flow direction (i.e., droplet front) experience higher surface vapour gradients, whereas larger concentrations are found in those where the normal vector is opposite to the flow (i.e. droplet rear). Note also that the increase of $We_{\mathcal{S}}$ corresponds to a shift of the most probable values of $\dot{m}_{\Gamma}$ and $\kappa_{\Gamma}$ towards higher values.}

\begin{figure*}
    \centering
    \includegraphics[width=12.60 cm, height=4.335 cm]{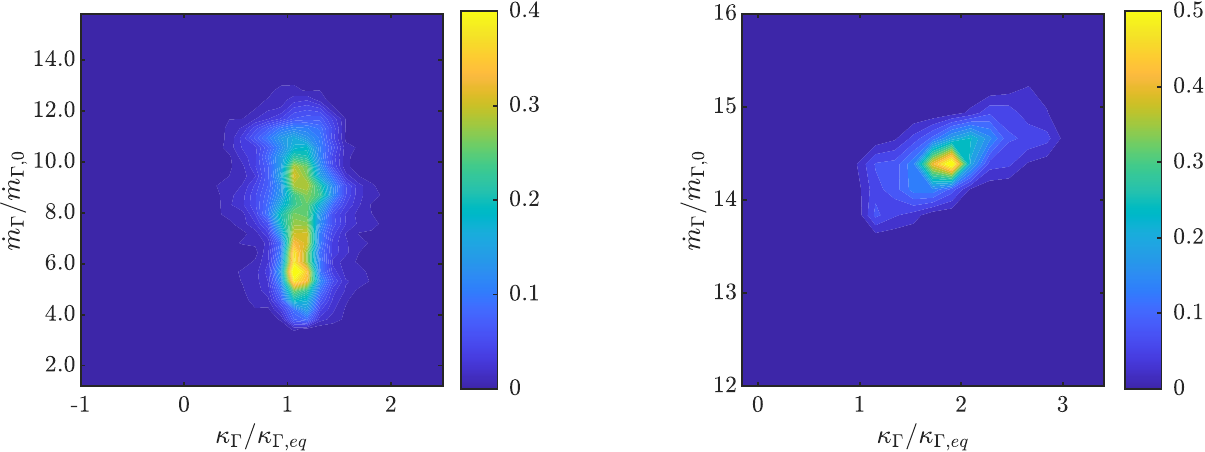}
    \put(-361,114){\small(\textit{a})}
    \put(-163,114){\small(\textit{b})}
    \caption{Joint p.d.f. of the dimensionless mass flux $\dot{m}_{\Gamma}/\dot{m}_{\Gamma,0}$ and the normalized interfacial curvature $\kappa_{\Gamma}/\kappa_{\Gamma,eq}$ for $Re_{\mathcal{S}}=6700$, $T_{g,0}/T_c=0.75$ and a) $We_{\mathcal{S}}=0.02$, b) $We_{\mathcal{S}}=0.10$.}
    \label{fig:jpdf_mk_hw_hr}
\end{figure*}

\begin{figure*}
    \centering
    \includegraphics[width=12.60 cm, height=4.335 cm]{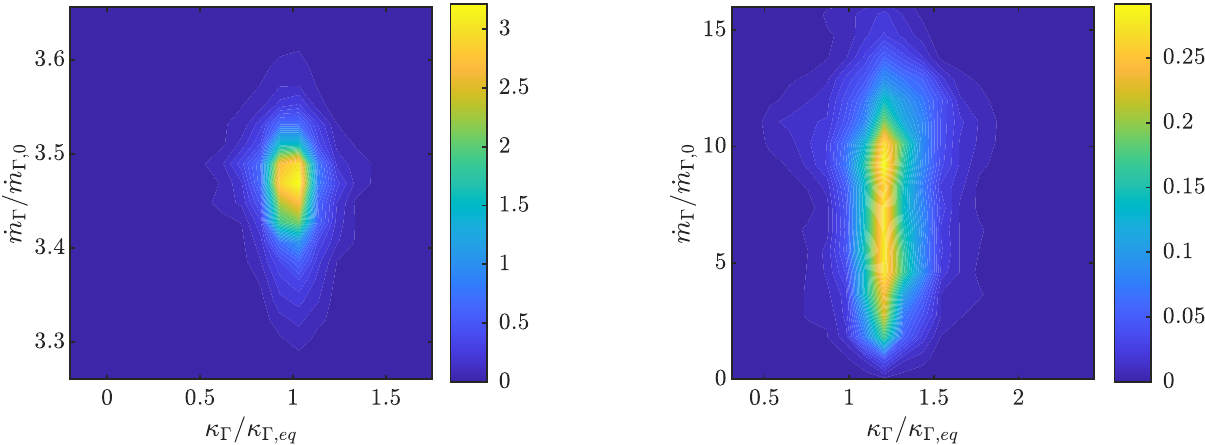}
    \put(-361,114){\small(\textit{a})}
    \put(-163,114){\small(\textit{b})}
    \caption{Joint p.d.f. of the dimensionless mass flux $\dot{m}_{\Gamma}/\dot{m}_{\Gamma,0}$ and the normalized interfacial curvature $\kappa_{\Gamma}/\kappa_{\Gamma,eq}$ for $Re_{\mathcal{S}}=2800$ for $We_{\mathcal{S}}=0.02$ and a) $T_{g,0}/T_c=1.50$ and b) $T_{g,0}/T_c=0.75$.}
    \label{fig:jpdf_mk_lr}
\end{figure*}
When the gas temperature is reduced (i.e., $T_{g,0}/T_c=0.75$), the correlation between $\dot{m}_{\Gamma}$ and $\kappa_{\Gamma}$ is still weak for $We_{\mathcal{S}}=0.02$, while for the case at $We_{\mathcal{S}}=0.10$ a higher interfacial curvature is associated with a higher interfacial mass flux. Also at lower temperatures, the most probable values are located at higher $\kappa_{\Gamma}$ when increasing the Weber number. Comparing with the results at higher temperatures,  figure~\ref{fig:jpdf_mk_lw_hr}, we note that $\dot{m}_{\Gamma}/\dot{m}_{\Gamma,0}$ is larger: turbulence enhances evaporation more efficiently at low temperature, when the evaporative time scale is longer and the ratio with the turbulent time scales decreases, i.e.\ $Da_v$ decreases.\\
%
%
%
Finally, we display in figure~\ref{fig:jpdf_mk_lr} the joint p.d.f. from the simulations at 
 lower Reynolds number, $Re_{\mathcal{S}}=2800$, for $T_{g,0}/T_c=0.75-1.5$ and $We_{\mathcal{S}}=0.02$. In this case,  the reduction of $T_{g,0}/T_c$ leads to more pronounced deformation, i.e.\ larger values of $\kappa_{\Gamma}/\kappa_{\Gamma,eq}$ (due to the slower evaporation and higher effective $We_{\mathcal{S}}$) and to higher values of $\dot{m}_{\Gamma}/\dot{m}_{\Gamma,0}$ (more pronounced effects of turbulence at lower absolute evaporation rates) and a broader distribution, again attributed to fluctuations induced by the turbulence. Comparing with the results at the same $We_{\mathcal{S}}$ and $T_{g,0}/T_c$ but $Re_{\mathcal{S}}=6700$ in panels $(a)$ of figures~\ref{fig:jpdf_mk_lw_hr} and~\ref{fig:jpdf_mk_hw_hr}, we see that, for both temperature ratios, the reduction in $Re_{\mathcal{S}}$ restricts the range of attained values for the curvature and the ratio $\dot{m}_{\Gamma}/\dot{m}_{\Gamma,0}$, confirming that the increase of deformation and evaporation rate is more pronounced for bigger droplets. 

\section{Conclusions}\label{sec:concl}
Fully resolved simulations of finite-size evaporating droplets are performed in homogeneous shear turbulence (HST) using a weakly compressible solver for evaporating flows. The new methodology, here described, combines two features. First, an improved version of the Adams-Bashforth method in order to address the limitations of the classical formulation, already highlighted in~\cite{kasbaoui2017algorithm}. The improvement allows us to match the single-phase analytical solution in the Rapid Distortion Limit and ensures a stable integration over long times. Second, a mathematical model with the details of the numerical implementation for a two-phase evaporating system composed of an incompressible liquid and a compressible gas phase. In order to remove the acoustic time-step restriction, compressible effects are here handled in the low-Mach number limit. This new methodology is applied to investigate the behaviour of finite-size evaporating droplets in HST when changing the gas temperature over the critical temperature $T_{g,0}/T_c=0.75, 1.00$ and $1.50$, the initial droplet diameter in terms of Kolmogorov scale, $d_0/\eta=10.5$ and $21.5$ and the surface tension, quantified by the shear-based Weber number $We_{\mathcal{S}}=0.02, 0.06$ and $0.10$. \\
%
%
%
First, using the data at $d_0/\eta= 21.5$ and $We_{\mathcal{S}}=0.02$, we study the differences when employing different thermodynamic models for the gas thermophysical properties. Three approaches are investigated: a constant property model where the gas properties are kept constant and initialized with the "1/3 rule" (CP) and two variable-properties approaches where either the gas density, VP$_{\rho}$ or all the gas properties are allowed to vary, VP$_a$. We find that the predictions by the CP and VP$_a$ models agree well, whereas the VP$_{\rho}$ model overpredicts the evaporation rate, especially at high temperature. This overestimation occurs since the local increase of the gas density (due to evaporative cooling) is captured by the $VP_{\rho}$ model, but the decrease of the diffusion coefficient with temperature, which slows down the evaporation, is not accounted. \nsc{Moreover, by extracting the Sherwood number for the three models at $T_{g,0}/T_c = 0.75$, $1.50$ and comparing it with the Fr{\"o}ssling/Ranz-Marshall correlation~\citep{ranz1952evaporation}, we show that the correlation provides an excellent estimation of $Sh$ at high temperature (conduction-dominated regime), whereas it substantially underestimates $Sh$ at lower temperature (convective-dominated regime), in agreement with recent experimental observations~\citep{mees2020statistical}}. \\
%
%
Next, reducing the ratio $d_0/\eta$ from 21.5 to 10.5 (obtained by reducing $Re_{\mathcal{S}}$ from 6700 to 2800), we show that the ratio between the actual evaporation rate, and the one computed in stagnant conditions, is always much higher than 1, while decreasing with $T_{g,0}/T_c$. Interestingly, this ratio does not approach unity at the highest temperature level, suggesting an evaporation enhancement due to turbulence also in these conditions. The variation of the droplet size, $d_0/\eta$, also affects the liquid temperature at the interface and the mean liquid temperature. For $T_{g,0}/T_c=1.0, 1.5$, the regime values of both quantities are larger for droplets of $d_0/\eta=21.5$, whereas the 
opposite is true  at $T_{g,0}/T_c=0.75$ when the regime values of the liquid temperature are lower for the largest droplets. This is explained as the result of two competing effects: cooling by evaporation and heating from the hot gas. \\
%
%
Finally, by varying $We_{\mathcal{S}}$ in the range $0.02$ and $0.10$, we observe an increase in the evaporation rate for higher $We_{\mathcal{S}}$ given the larger surface area available for mass transfer. At fixed $We_{\mathcal{S}}$, this increase is more pronounced at $T_{g,0}/T_c=0.75$ due to a slower evaporation rate and to higher deformation (larger effective Weber number based on the instantaneous droplet diameter). By computing the joint p.d.f. of the interfacial mass flux and curvature, a weak correlation between the two is observed at high temperature regardless of the Weber number, whereas a positive correlation is recovered at $T_{g,0}/T_c=0.75$ and $We_{\mathcal{S}}=0.10$, consistently with the theoretical prediction in laminar environments~\citep{tonini2013exact} and the experimental observations for sessile droplets~\citep{saenz2017dynamics}. \\
%
%
Note that in all the cases, the initial liquid volume fraction has been kept small (i.e., $\alpha_0\approx 0.14$ $\%$), the droplets do not interact with each other and, as expected, no significant deviations from the $d^2$-law have been observed. To investigate such deviations and more in general the collective dynamics of a dense suspension of evaporating droplets we are currently exploring the same configurations presented in this work at higher volume fractions. 
\section*{Acknowledgements}
N.S. and L.B. acknowledge the support from the Swedish Research Council via the multidisciplinary research environment INTERFACE, Hybrid multiscale modelling of transport phenomena for energy efficient processes and the Grant No.\ 2016-06119. \nsc{M.E.R. was supported by the JSPS KAKENHI Grant Number JP20K22402}. The computer time was provided by SNIC (Swedish National Infrastructure for Computing) and by the National Infrastructure for High Performance Computing and Data Storage in Norway (project no. NN9561K). Francesco Picano and Marco Crialesi Esposito are thanked for the useful discussions and, the latter, also for providing the droplet tagging algorithm. \nsc{Karim Khamis is acknowledged for pointing out to some typos in the dimensionless formulation of the mathematical model}.


\appendix

\section{Equations of state}\label{sec:eqn_state}
In this Appendix, we report the equations of state employed for the gas density, the gas thermal diffusivity, the gas viscosity, thermal conductivity, heat capacity at constant pressure and the Span-Wagner equation of state. 

\subsection{Gas density}
The gas density variations as a function of $p_{th}$, $T_g$ and $Y_{l}^v$ follow the ideal gas model
\begin{equation}
	\rho_g = \Pi_{p,2}\dfrac{p_{th}\overline{M}_{m,av}}{T_g}\mathrm{,}
  \label{gas_density}
\end{equation}
where $\overline{M}_{m,av}$ is the mixture molar mass computed using an harmonic average between the gas and the liquid molar mass and $\Pi_{p,2}=(p_{th,r}M_g)/(R_uT_{g,r})$ where $p_{th,r}$ and $T_{g,r}$ are the reference thermodynamic pressure and gas temperature, taken equal to the corresponding initial value and $R_u$ is the universal gas constant.

\subsection{Liquid-gas diffusion coefficient}
The liquid gas diffusion coefficient is computed using the relation~\citep{reid1987properties}:
\begin{equation}
	D_{lg} = \dfrac{T_g^{3/2}}{p_{th}}\mathrm{,}
\end{equation}
where $T_g$ is the gas temperature.

\subsection{Gas viscosity}
The gas viscosity varies with the temperature according to the simplified Sutherland's law~\citep{reid1987properties}:
\begin{equation}
	\mu_g = T_{g}^{2/3}\mathrm{.}
\end{equation}

\subsection{Specific heat capacity at constant pressure}
The specific heat capacity at constant pressure is evaluated as a function of the temperature using the virial polynomials~\citep{reid1987properties}:
\begin{equation}
	c_{p} = A_1+A_2T_g+A_3T_g^2+A_4T_g^3+A_5T_g^4\mathrm{,}
\end{equation}
where the coefficients $A_1=1.012$, $A_2=0.0553$, $A_3=0.006$, $A_4=2\cdot 10^{-3}$ and $A_5=5\cdot 10^{-4}$ are adequate to estimate the heat capacity at constant pressure for dry air over a wide range of temperatures.

\subsection{Gas thermal conductivity}
As for the gas viscosity, the gas thermal conductivity is  function of the temperature only and computed as \citep{reid1987properties}:
\begin{equation}
	k_g = \dfrac{\mu_gc_{pg}}{Pr}\mathrm{,}
\end{equation}
where $Pr$ is the Prandtl number.

\subsection{Span–Wagner equation of state}
In order to compute the value of $Y_{l}^{v}$ at the interface, see section~\ref{sec:vap_mass}, we assume that the gas mixture is ideal and composed of ideal components. Hence, $Y_{l,\Gamma}^{v}$ can be computed using the Rault's law,
\begin{equation}
  Y_{l,\Gamma}^{v} = \dfrac{\dfrac{p_{s,\Gamma}}{p_{th}}}{\dfrac{p_{s,\Gamma}}{p_{th}}+\left(1-\dfrac{p_{s,\Gamma}}{p_{th}}\right)\dfrac{1}{\lambda_M}}\mathrm{,}
  \label{eqn:sp_wg_1}
\end{equation}
where $\lambda_M=M_l/M_g$ is the molar mass ratio, while $p_{th}$ and $p_{s,\Gamma}$ are the thermodynamic pressure and the partial pressure of the vaporized species at the interface. Since the interface is assumed at saturation, $p_{s,\Gamma}$ is computed as a function of the liquid temperature at the interface. For most of the substances, Span-Wagner equation of state represents a valid approximation over a wide range of thermodynamic pressures and temperatures~\citep{span1996new}:
\begin{equation}
	\dfrac{p_{s,\Gamma}}{p_{th}} = \dfrac{\Pi_{p,3}}{p_{th}}\exp\left[\dfrac{B_1\eta_{sw}+B_2\eta_{sw}^{1.5}+B_3\eta_{sw}^{2.5}+B_4\eta_{sw}^5}{1-\eta_{sw}}\right]\mathrm{.}
  \label{eqn:sp_wg_2}
\end{equation}
where $\eta_{sw}=1-T_{\Gamma}(T_{g,0}/T_c)$ and $\Pi_{p,3}=p_c/p_{th,r}$, with $T_{g,0}$, $T_c$ and $p_c$ the initial gas temperature, the critical temperature and critical pressure, respectively. The coefficients $B_{i=1,4}=\{-7.32714,1.82365,-2.272744,-2.711929\}$ are experimentally determined and correspond in the current study to those of pentane; see~\cite{span1996new}. Note that the use of Span-Wagner model is also convenient since it does not involve the saturation temperature (like the Clausius–Clapeyron relation), which in a weakly compressible phase-changing system is not a constant anymore, but should be computed as a function of the time-varying thermodynamic pressure. \\
\revC{As a final remark, to justify the assumption of incompressible liquid with constant properties, we evaluate a posteriori 
  the variations of $\rho_l$, $c_{pl}$, $k_l$ and $\mu_l$
 given the maximum variation of liquid temperature discussed in section~\ref{sec:re_effect}. Using the data reported in~\cite{reid1987properties} and taking as reference the cases at $T_{g,0}/T_c=1.5$ (when the liquid temperature variation is more pronounced), these amount to: $\Delta\rho_l/\rho_{l,0}=6$ $\%$, $\Delta\mu_l/\mu_{l,0}=8$ $\%$, $\Delta c_{pl}/c_{pl,0}=4$ $\%$, $\Delta k_l /k_{l,0}=5$ $\%$}.

\section{Derivation of the velocity divergence}\label{sec:vel_div}

The system composed of equations~\eqref{eqn:mom1},~\eqref{eqn:vap1},\eqref{eqn:int_mfx} and~\eqref{eqn:tmp1} reported in section~\ref{sec:gov_eqn} is not closed and an additional equation should be included. To derive it, we adapt the approach proposed in~\cite{majda1985derivation} in the context of reacting flows and recently employed for phase change in~\cite{dodd2021analysis}. Nevertheless, since a detailed derivation is missing in literature, it is provided here for completeness. The idea is to compute the missing relation from the divergence of the velocity field, $\mathbf{u}$, which can be computed as $H\mathbf{u}_l+(1-H)\mathbf{u}_g$. Taking the divergence yields
\begin{equation}
	\nabla\cdot\mathbf{u}=(\mathbf{u}_g-\mathbf{u}_l)\cdot\nabla H+(1-H)\nabla\cdot\mathbf{u}_g+H\nabla\cdot\mathbf{u}_l\mathrm{.}
	\label{div_u}
\end{equation}
The first term on the right-hand side represents the volume variation at the interface due to phase change, while the second and the third come from the density variations in the gas and in the liquid phases. We start with the phase-change contribution and consider a reference frame moving with the interface. Accordingly, we decompose the vector $\mathbf{u}_g-\mathbf{u}_l$ along the normal $\mathbf{n}_{\Gamma}$, tangential $\mathbf{t}_{\Gamma}$ and bi-normal directions $\mathbf{n}_{\Gamma}$:
\begin{equation}
	\mathbf{u}_g-\mathbf{u}_l=(u_{g,n}-u_{l,n})\mathbf{n}_{\Gamma}+(u_{g,t}-u_{l,t})\mathbf{t}_{\Gamma}+(u_{g,b}-u_{l,b})\mathbf{b}_{\Gamma}.
	\label{total_jump}
\end{equation}
If the interface has zero thickness, the mass balance imposes that the velocity is continuous along the tangential and bi-normal direction (i.e., $u_{g,t}=u_{l,t}$ and $u_{g,b}=u_{l,b}$). Along the normal direction, conversely, the velocity has a discontinuity proportional to the mass flux $\dot{m}_{\Gamma}$ between the two phases, i.e.,
\begin{equation}
  \begin{aligned}
	  &\dot{m}_{\Gamma} = \rho_{g,\Gamma}(u_{g,n}-u_{\Gamma,n})\mathrm{,} \\
	  &\dot{m}_{\Gamma} = \rho_{l,\Gamma}(u_{l,n}-u_{\Gamma,n})\mathrm{,}
  \end{aligned}
\end{equation}
where $u_{\Gamma,n}$ represents the normal component of the interface velocity and $\rho_{i=l,g,\Gamma}$ the phase density at the interface location. Combining the previous two equations yields
\begin{equation}
	u_{g,n}-u_{l,n}=\dot{m}_{\Gamma}\left(\dfrac{1}{\rho_{g,\Gamma}}-\dfrac{1}{\rho_{l,\Gamma}}\right)\mathrm{.}
	\label{jump_n}
\end{equation}
Inserting equations~\eqref{total_jump} and~\eqref{jump_n} into equation~\eqref{div_u}, we obtain 
\begin{equation}
	(\mathbf{u}_g-\mathbf{u}_l)\cdot\nabla H=\dot{m}_{\Gamma}\left(\dfrac{1}{\rho_{g,\Gamma}}-\dfrac{1}{\rho_{l,\Gamma}}\right)\delta_{\Gamma}\mathrm{,}
	\label{div_u1}
\end{equation}
where $\delta_{\Gamma}=\nabla H\cdot\mathbf{n}_{\Gamma}$. \\
The second term of equation~\eqref{div_u} is computed from the continuity in the gas region. If chemical reactions are absent, this reads as
\begin{equation}
	\nabla\cdot\mathbf{u}_g = -\dfrac{1}{\rho_g}\dfrac{D\rho_g}{Dt}\mathrm{.}
\end{equation}
Using the equation of state for the gas density~\eqref{gas_density}, the total derivative of $\rho_g$ can be expanded as follows:
\begin{equation}
	\dfrac{1}{\rho_g}\dfrac{D\rho_g}{Dt} = \dfrac{1}{p_{th}}\dfrac{Dp_{th}}{Dt}-\dfrac{1}{T}\dfrac{DT}{Dt}-\overline{M}_{m,av}\dfrac{D}{Dt}\left(\dfrac{1}{\overline{M}_{m,av}}\right)\mathrm{.}
     \label{drhogdt}
\end{equation}
Combining eq.~\eqref{div_u1} and~\eqref{drhogdt} finally provides the velocity divergence in the gas region~\cite{dodd2021analysis},
\begin{equation}
	\nabla\cdot\mathbf{u}_g=-\dfrac{1}{p_{th}}\dfrac{dp_{th}}{dt}+\dfrac{1}{T}\dfrac{DT}{Dt}+\overline{M}_{m,av}\dfrac{D}{Dt}\left(\dfrac{1}{\overline{M}_{m,av}}\right)\mathrm{.}
  \label{divug} 
\end{equation}
Note that, given the low-Mach assumption, the total derivative of $p_{th}$ in~\eqref{divug} has been replaced by the time derivative of $p_{th}$. 
A similar strategy can be employed to compute $\nabla\cdot\mathbf{u}_l$; however, we consider here  constant and uniform liquid density and therefore, $\nabla\cdot\mathbf{u}_l=0$ in eq.~\eqref{div_u} and  $\rho_{l,\Gamma}=\rho_l$. \par
Differently from what it is commonly done in literature~\citep{motheau2016high}, we found preferable to replace the total derivative of the molar mass and of the temperature with the corresponding spatial derivatives. By doing so, this solution completely removes the time discretization errors when computing the terms in~\eqref{divug}. For the contribution due to composition, we employ equation~\eqref{eqn:vap1} and we treat the mixture of inert gas and vapour as ideal, so that the mixture molar mass is given by the harmonic average of the molar mass of each component~\citep{reid1987properties}. Accordingly, using the relation $Y_g=Y_N=1-\sum_{j=1}^{N-1}Y_{l,j}^v$ the third term on the right-hand side of equation~\eqref{divug} becomes
\begin{equation}
	\overline{M}_{m,av}\dfrac{D}{Dt}\left(\dfrac{1}{\overline{M}_{m,av}}\right) = \overline{M}_{m,av}\sum_{j=1}^{N-1}\left(\dfrac{1}{\lambda_{M,j}}-1\right)\dfrac{DY_{l,j}^v}{Dt}\mathrm{.}
  \label{eqn:vap_mass}
\end{equation}
where $\lambda_{M,j}=M_{l,j}/M_g$. For the temperature $T$, we start from the generic enthalpy equation for the gas phase: 
\begin{equation}
	\rho\dfrac{Dh}{Dt} = \dfrac{1}{RePr}\nabla\cdot(k\nabla T) + \Pi_{p,1}\dfrac{dp_{th}}{dt}+\dfrac{1}{ReSc}\sum_{j=1}^N h_{j}^v\nabla\cdot\left(\rho_gD_{lg}\nabla Y_{j}\right)\mathrm{.} 
  \label{eqn:enthalpy} 
\end{equation}
In a weakly compressible system, being the enthalpy of the gas phase function of temperature, thermodynamic pressure and vapour composition, the left-hand side of~\eqref{eqn:enthalpy} can be expanded as
\begin{equation}
  \begin{split}
	  \rho\dfrac{Dh}{Dt} &=\rho\left(\dfrac{\partial h}{\partial T}\right)_{Y_{l,j}^v,p_{th}}\dfrac{DT}{Dt} + \rho\left(\dfrac{\partial h}{\partial p_{th}}\right)_{Y_{l,j}^v,T}\dfrac{dp_{th}}{dt}+\rho\sum_{j=1}^{N-1}\left(\dfrac{\partial h}{\partial Y_{l,j}^v}\right)_{T,Y_{i\neq j}}\dfrac{DY_{l,j}^v}{Dt}\mathrm{,} \\
		       &=\rho c_p\dfrac{DT}{Dt} + \rho(1-\beta T)\dfrac{dp_{th}}{dt} + \rho\sum_{j=1}^{N-1}\left(\dfrac{\partial h}{\partial Y_{l,j}^v}\right)_{T,p_{th},Y_{l,i\neq j}^v}\dfrac{DY_{l,j}^v}{Dt}\mathrm{.} \\
  \end{split}
  \label{eqn:ent_split}
\end{equation}
Note that for an ideal gas, the isothermal compressibility coefficient $\beta=1/T$ and, thus, $(1-\beta T)=0$, which can be omitted. Considering the term varying with the composition $Y_{l,j}^v$, this can be recast as \cite[see][for details]{lupo2019immersed}:
\begin{equation}
	\sum_{j=1}^{N-1}\left(\dfrac{\partial h}{\partial Y_{l,j}^v}\right)_{T,p_{th},Y_{l,i\neq j}^v}\dfrac{DY_{l,j}^v}{Dt} = \sum_{j=1}^{N-1}\left(h_{v,j}-h_g\right)\dfrac{DY_{l,j}^v}{Dt}\mathrm{.}
  \label{eqn:lupo_dev}
\end{equation}
Inserting equations~\eqref{eqn:lupo_dev} and~\eqref{eqn:ent_split} into equation~\eqref{eqn:enthalpy} and using~\eqref{eqn:vap1}, we get
\begin{equation}
	\rho c_p\dfrac{DT}{Dt} = \dfrac{1}{RePr}\nabla\cdot(k\nabla T) + \left(\Pi_{p,1}\dfrac{dp_{th}}{dt}+\dfrac{\rho_g}{ReSc}\sum_{j=1}^{N}D_{lg,j}\nabla h_{j}\cdot\nabla Y_{j}\right)(1-H)\mathrm{.} 
  \label{eqn:tmp}
\end{equation}
Note that equation~\eqref{eqn:tmp} is the same as~\eqref{eqn:tmp1} except for the source term due to phase change which acts at the interface location. Finally, by combining~\eqref{eqn:tmp},~\eqref{eqn:vap_mass}~\eqref{divug} in expression~\eqref{div_u}, the velocity divergence $\mathbf{u}$ can be expressed as
\begin{align}
	\nabla\cdot\mathbf{u} &= \,\, \underbrace{\dot{m}_{\Gamma}\left(\dfrac{1}{\rho_{g,\Gamma}}-\dfrac{1}{\lambda_{\rho}}\right)\delta_{\Gamma}}_{f_{\Gamma}(\mathbf{x}_{\Gamma},t)}\mathrm{,} \nonumber \\ & +\dfrac{1}{p_{th}}\underbrace{\dfrac{1}{ReSc}\dfrac{\overline{M}_{m,av}}{\rho_g}\sum_{j=1}^{N-1}\left(\dfrac{1}{\lambda_{M,j}}-1\right)\nabla\cdot(\rho_gD_{lg,j}\nabla Y_{l,j}^v)}_{f_Y(\mathbf{x},t)}(1-H)\mathrm{,} \nonumber \\ &+ \dfrac{1}{p_{th}}\underbrace{\dfrac{1}{Re}\dfrac{\Pi_{p,1}}{c_p\overline{M}_{m,av}}\left[\dfrac{1}{Pr}\nabla\cdot(k\nabla T)+\dfrac{\rho_g}{Sc}\sum_{j=1}^ND_{lg,j}\nabla h_j\cdot\nabla Y_j\right]}_{f_T(\mathbf{x},t)}(1-H) \nonumber\mathrm{,} \\ &- \left(1-\dfrac{\Pi_{p,1}}{c_p\overline{M}_{m,av}}\right)\dfrac{1}{p_{th}}\dfrac{dp_{th}}{dt}(1-H)\mathrm{,}
  \label{eqn:veldiv}
\end{align}
where $f_{\Gamma}$, $f_Y$ and $f_T$ are functions representing the different contributions to the total velocity divergence, i.e., phase change, ($f_{\Gamma}$), change of the gas density due to composition ($f_Y$) and due to temperature ($f_T$). \par 
The last step is to derive an expression for the thermodynamic pressure, $p_{th}$. For an open domain, $p_{th}$ is constant, whereas for a closed or triperiodic domain it can be obtained by imposing the volume conservation on the global domain $V$. This can formally expressed as a constrain on the velocity divergence
\begin{equation}
	\int_V\nabla\cdot\mathbf{u} \,\,dV = 0\mathrm{.}
  \label{eqn:kinematic}
\end{equation}
By employing~\eqref{eqn:kinematic}, we can easily rearrange the rate of change of $p_{th}$ as
\begin{equation}
	\dfrac{1}{p_{th}}\dfrac{dp_{th}}{dt}\int_{V_g}\left(1-\dfrac{\Pi_{p,1}}{c_p\overline{M}_{m,av}}\right)dV_g = \int_V\left[f_{\Gamma}(\mathbf{x}_{\Gamma},t)+\dfrac{(f_T(\mathbf{x},t)+f_Y(\mathbf{x},t))}{p_{th}}(1-H)\right] dV\mathrm{.}
  \label{eqn:veldiv}
\end{equation}

\section{Verification/validation of the low-Mach solver with phase change}\label{sec:val_lm}
\revA{In this appendix, we provide a verification and two validation cases of the weakly compressible code employed to perform the numerical simulations in the present work.}

\subsection{Droplet evaporation due to a prescribed, constant mass flux}
\begin{figure*}
    \centering
    \includegraphics[width=6.5 cm,height=5.2 cm]{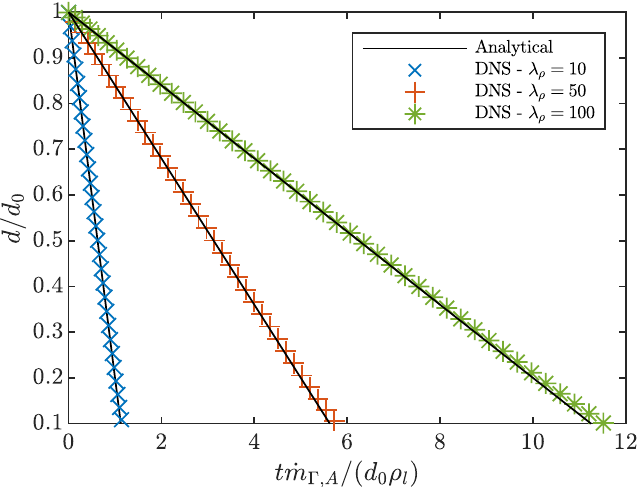}
    \includegraphics[width=6.5 cm,height=5.2 cm]{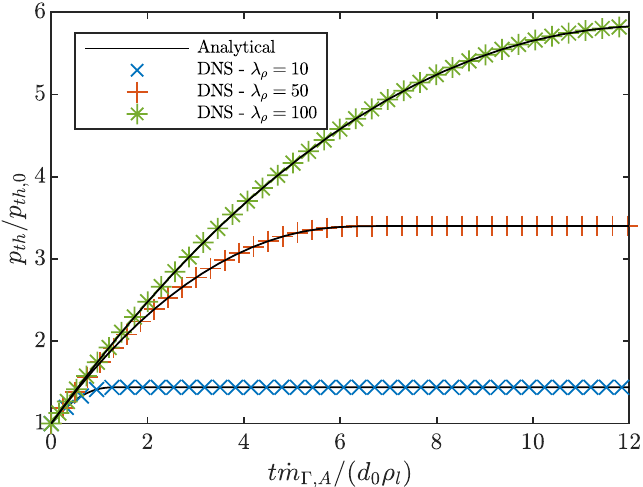}
    \put(-370,138){\small(\textit{a})}
    \put(-187,138){\small(\textit{b})}
    \caption{\textit{a}): Temporal evolution of the instantaneous droplet diameter over the initial value, $d/d_0$, \textit{b}): Temporal evolution of the thermodynamic pressure over the initial value, $p_{th}/p_{th,0}$.}
    \label{fig:validation1}
\end{figure*}

The set-up of this verification case consists of a two-dimensional circular droplet with initial diameter $d_0$, which evaporates due to a prescribed, constant mass flux $\dot{m}_{\Gamma,A}$. In this case, the momentum equation is decoupled from the transport of energy and vapour mass fraction equations and it is straightforward to show that the droplet diameter evolves as:
\begin{equation}
	\dfrac{d(t)}{d_0} = 1-\left(\dfrac{2\dot{m}_{\Gamma,A}}{d_0\rho_l}\right)t\mathrm{.}
  \label{eqn:dt}
\end{equation}
In our previous work \citep{scapin2020volume}, we reproduce this test case using a zero-pressure outflow boundary condition, whereas here we prescribe periodic boundary conditions. Accordingly, as evaporation starts, the thermodynamic pressure builds up and eventually reaches a stationary value when the droplet is completely evaporated. \nsc{By setting $f_Y$ and $f_T$ to zero in equation~\eqref{eqn:veldiv} and after some manipulations a nonlinear ordinary differential equation can be derived for $p_{th}$,}
\begin{equation}
	\dfrac{dp_{th}(t)}{dt} - \dfrac{42\dot{m}_{\Gamma,A}\pi d^2(t)}{30G_{tot}-5\rho_l\pi d^3(t)}\left(1-\dfrac{p_{th}(t)}{\rho_l(R_u/M_g)T_0}\right)p_{th}(t)=0\mathrm{,}
  \label{eqn:pth_ex}
\end{equation}
where $d(t)$ is computed using equation~\eqref{eqn:dt}, $G_{tot}$ is the mass inside the system, taken equal to its initial value (i.e., $G_{tot}=\rho_{g,0}V_{g,0}+\rho_lV_{l,0}$), $T_0$ and $(R_u/M_g)$ are chosen so that the group $p_{th,0}/(\rho_{g,0}(R_u/M_g)T_0)=1$. 
Equation~\eqref{eqn:pth_ex} is here solved with the fourth-order Runge-Kutta scheme. The test has been repeated for three density ratios $\lambda_{\rho}=10-50-100$, with the remaining dimensionless physical parameters $Re=25$, $We=0.10$, $\lambda_{\mu}=50$; gravity is absent. The governing equations are solved in a square domain $[-2d_0;2d_0]^2$, discretized with $128\times 128$ grid points. The droplet is held at the centre of the domain. Results are reported in figure~\ref{fig:validation1}: an excellent agreement between the numerical and the analytical solutions is found for all the cases.

\revA{
\subsection{Hexadecane static droplet evaporation in a hot gas}\label{subsec:hexadecane}
As a validation case, we reproduce the numerical test proposed in~\cite{ni2021numerical} consisting of a static hexadecane droplet of initial diameter $d_0=550$ $\mu m$, which evaporates in dry air at $T_g=673$ K. For this test case, we set $Re=25$, $We=0.1$, $Pr=Sc=1$, $\lambda_{\rho}=770$, $\lambda_{\mu}=202$, $\lambda_k=20$, $\lambda_{c_p}=2.1$. Zero-pressure outflow boundary conditions are prescribed (therefore, $p_{th}=p_{th,0}$) and the assumption of constant liquid bulk density is relaxed \cite[see][for more details]{ni2021numerical}. The whole set of governing equations are solved in a square domain $[-10d_0;10d_0]^2$ using $256\times 256$ grid points. Once evaporation starts, the liquid droplet undergoes a local initial expansion (i.e., $(d/d_0)^2>1$) until $t/d_0^2\approx 2$, after which the $d^2$ regime is approached (see figure~\ref{fig:validation23}a). Overall, good agreement between our simulations and the reference data is observed, confirming once more the validity of our numerical algorithm. 


\subsection{Single droplet evaporating in homogeneous isotropic turbulence}\label{subsec:heptane}
As a final validation case, we reproduce the experimental results in~\cite{verwey2018fuel}, reproduced also in~\cite{dodd2021analysis}, consisting of a single n-heptane droplet of initial diameter $d_0=200$ $\mu m$, which evaporates in nitrogen at $T_g=348$ K and $p_{th,0}=10$ bar. For this test case, we consider homogeneous isotropic  turbulence sustained by an artificial forcing~\citep{podvigina1994non} at a Reynolds number based on the Taylor's microscale $Re_{\lambda}=u_{rms}'\lambda/\nu_{g,r}\approx 32$. The initial droplet Weber number $We_{rms}=\rho_{g,r}u_{rms}'^2d_0/\sigma$ is set equal to $0.0044$, $Pr=0.715$, $Sc=2.81$, $Ste=1.05$, $\lambda_{\rho}=25.27$, $\lambda_{\mu}=11.82$, $\lambda_k=3.72$ and $\lambda_{c_p}=2.45$. The whole set of governing equations are solved in a cubic domain $[-16d_0;16d_0]^3$ using $768^3$ grid points corresponding  to about $48$ grid points per initial diameter. Figure~\ref{fig:validation23}b reports $(d/d_0)^2$ as a function of $t/d_0^2$ for the both the experimental and the numerical results. A good agreement is observed with a deviation of evaporation rate $K_{DNS}$ of less then $7$ $\%$ with respect to the experimental data, which is considered a satisfactory result and within the uncertainty of the measurement~\citep{verwey2018fuel}.

\begin{figure*}
    \centering
    \includegraphics[width=6.5 cm,height=5.2 cm]{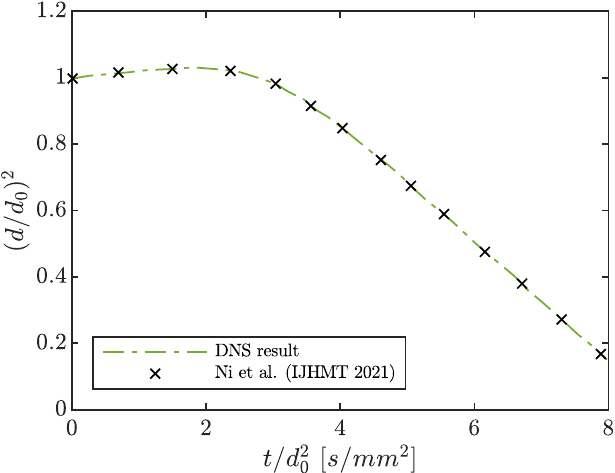}
    \includegraphics[width=6.5 cm,height=5.2 cm]{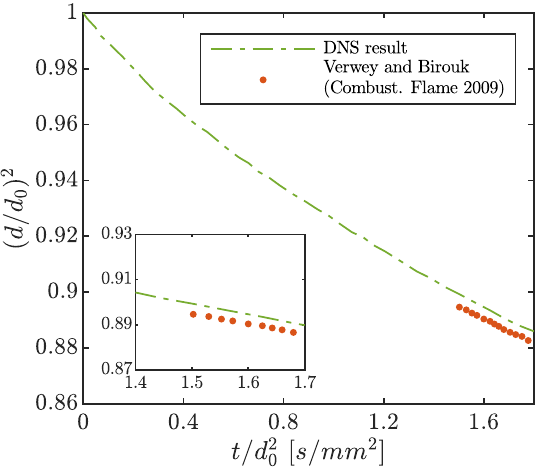}
    \put(-374,138){\small(\textit{a})}
    \put(-185,138){\small(\textit{b})}
    \caption{Temporal evolution of the instantaneous droplet diameter $(d/d_0)^2$ over the initial value for the \textit{a}): validation case described in~\ref{subsec:hexadecane} and the \textit{b}):~\ref{subsec:heptane}.}
    \label{fig:validation23}
\end{figure*}
}

\section{Improved Adams-Bashforth scheme for HST simulations}\label{sec:imp_ab2}
\cite{kasbaoui2017algorithm} show that that the classical Adams-Bashforth scheme (AB2) employed in~\cite{gerz1989direct} is not suitable for the direct numerical simulations of homogeneous shear turbulence. Using as a benchmark the Kelvin modes derived in the framework of the Rapid Distortion Theory (RDT)~\citep{maxey1982distortion,isaza2009asymptotic}, they show that the method fails to reproduce the analytical solution with an unbounded growth of the error. In this appendix, we propose a modification of the classical Adams-Bashforth scheme able to reproduce the Kelvin modes and, more generally, stable and accurate simulations. \par 

\subsection{Proof the numerical stability}
We follow the approach proposed in~\cite{kasbaoui2017algorithm} and compute the time discretization error introduced by the classical Adams-Bashforth scheme using the simplified equation
\begin{equation}
	\dfrac{\partial u}{\partial t} + \mathcal{S}z\dfrac{\partial u}{\partial x} = Au\mathrm{,}
  \label{eqn:toy_hst}
\end{equation}
where $A$ is a positive constant and $\mathcal{S}$ is the applied shear. As 
noted in \cite{kasbaoui2017algorithm}, equation~\eqref{eqn:toy_hst} allows us to remove the complications due to the pressure gradient while preserving some important key features of homogeneously sheared flows, i.e.\ the exponential growth of the turbulent kinetic energy~\citep{maxey1982distortion}. The solution of equation~\eqref{eqn:toy_hst}, $u=\hat{u}(t)\exp(i\mathbf{k}\cdot\mathbf{x})$, needs to satisfy two conditions for the amplitude $\hat{u}(t)$ and for the wave vector $\mathbf{k}$,
\begin{subequations}
	\begin{align}
		&\dfrac{d\hat{u}}{dt} = A\hat{u}\mathrm{,}
		\label{eqn:first_cond} \\
		&\dfrac{d\mathbf{k}}{dt}\cdot\mathbf{x} = -\mathcal{S}z\mathbf{k}\cdot\mathbf{e}_x\mathrm{.}	\label{eqn:sec_cond}
	\end{align}
\end{subequations}
Assuming 
that the discretization error evolves as a wave of amplitude $\hat{\varepsilon}$ and wavevector $\mathbf{k}$, at the $n$th time step 
\begin{equation}
	\varepsilon^n=\hat{\varepsilon}^n\exp(i\mathbf{k}^n\cdot\mathbf{x})\mathrm{.}
\end{equation}
Employing the classical Adams-Bashforth integration scheme, we obtain a first intermediate error neglecting the mean shear contribution,
\begin{equation}
	\overline{\varepsilon}^{n+1}=\varepsilon^n+A\Delta t^{n+1}\left[\gamma_1\mathbf{RU}(\hat{\varepsilon}^n,\mathbf{k}^n)-\gamma_2\mathbf{RU}(\hat{\varepsilon}^{n-1},\mathbf{k}^{n-1})\right]\mathrm{,}
  \label{err} 
\end{equation}
where $\mathbf{RU}$ represents the discrete operator for the spatial terms in the governing equation of the error and $\gamma_1=(1+0.5\Delta t^{n+1}/\Delta t^n)$ and $\gamma_2=0.5\Delta t^{n+1}/\Delta t^n$ are the two coefficients of the Adams-Bashforth scheme for a variable time-step size. Note again that, at this stage, the mean shear contribution is not included in $\mathbf{RU}$. \\
Following the approach proposed by~\cite{gerz1989direct}, we set $\mathbf{RU}(\hat{\epsilon}^n,\mathbf{k}^n)=\hat{\varepsilon}^n\exp(i\mathbf{k}^n\cdot\mathbf{x})$ and $\mathbf{RU}(\hat{\epsilon}^{n-1},\mathbf{k}^{n-1})=\hat{\varepsilon}^{n-1}\exp(i\mathbf{k}^{n-1}\cdot\mathbf{x})$. Moreover, by using the identity $\exp(i\mathbf{k}^{n-1}\cdot\mathbf{x})=\exp(-i\Delta\mathbf{k}\cdot\mathbf{x})\exp(i\mathbf{k}^n\cdot\mathbf{x})$, equation~\eqref{err} becomes
\begin{equation}
  \overline{\varepsilon}^{n+1}=\left\{\varepsilon^n+A\Delta t^{n+1}\left[\gamma_1\hat{\varepsilon}^n-\gamma_2\hat{\varepsilon}^{n-1}\exp(-i\Delta\mathbf{k}\cdot\mathbf{x})\right]\right\}\exp(i\mathbf{k}^n\cdot\mathbf{x}),
  \label{err_n0}
\end{equation}
where $\Delta\mathbf{k}=\mathbf{k}^n-\mathbf{k}^{n-1}$. Next, the mean shear contribution is included and this results in an error at time $n+1$, 
\begin{multline}
	\varepsilon^{n+1}=\overbrace{\left\{\varepsilon^n+A\Delta t^{n+1}\left[\gamma_1\hat{\varepsilon}^n-\gamma_2\hat{\varepsilon}^{n-1}\exp(-i\Delta\mathbf{k}\cdot(\mathbf{x}-\Delta t^{n+1}\mathcal{S}z\mathbf{e}_x))\right]\right\}}^{\hat{\varepsilon}^{n+1}}\cdot \\\exp[\underbrace{i\mathbf{k}^n\cdot(\mathbf{x}-\Delta t^{n+1}\mathcal{S}z\mathbf{e}_x)}_{i\mathbf{k}^{n+1}}]\mathrm{.}
  \label{err_n1a}
\end{multline}
If we analyse equation~\eqref{err_n1a}, we see that the new wave vector $\mathbf{k}^{n+1}$ satisfies the first condition, i.e., eqn~\eqref{eqn:first_cond}. In fact:
\begin{equation}
	\mathbf{k}^n\cdot(\mathbf{x}-\Delta t^{n+1}\mathcal{S}z\mathbf{e}_x) = \mathbf{k}^{n+1}\cdot\mathbf{x}\rightarrow \dfrac{\mathbf{k}^{n+1}-\mathbf{k}^n}{\Delta t^{n+1}}\cdot\mathbf{x} = -\mathcal{S}z\mathbf{k}^n\cdot\mathbf{e}_x\mathrm{.}
\end{equation}
Conversely, the amplitude at the new time level, $\hat{\varepsilon}^{n+1}$, does not satisfy the condition~\eqref{eqn:sec_cond} since it contains a spatial-dependent term, $\Delta\mathbf{k}$, which is caused by the difference in orientation between the wave vector at the current and  old time level, $n$ and $n-1$. \par
The approach proposed in this work improves the method by \cite{gerz1989direct} by removing the spatial-dependent term, $\Delta\mathbf{k}$. To this end, the third term on the right-hand side of equation~\eqref{err} is first shear mapped to the current time level,
\begin{equation}
	\hat{\varepsilon}^{n-1}\exp(i\mathbf{k}^{n-1}\cdot(\mathbf{x}-\Delta t^{n}\mathcal{S}z\mathbf{e}_x))=\hat{\varepsilon}^{n-1}\exp(i\mathbf{k}^n\cdot\mathbf{x})\mathrm{.}
	\label{err_rosti}
\end{equation}
Equation~\eqref{err_rosti} is key to obtain two wave errors of different amplitudes (i.e., $\hat{\varepsilon}^n$ and $\hat{\varepsilon}^{n-1}$) but equal wave vector, i.e., $\mathbf{k}^n$. Inserting equation~\eqref{err_rosti} in equation~\eqref{err} gives
\begin{equation}
	\overline{\varepsilon}^{n+1}=\left\{\varepsilon^n+A\Delta t^{n+1}\left[\gamma_1\hat{\varepsilon}^n-\gamma_2\hat{\varepsilon}^{n-1}\right]\right\}\exp(i\mathbf{k}^n\cdot\mathbf{x})\mathrm{.}
  \label{err_n1}
\end{equation}
Next, the shear-mapping step is applied to equation~\eqref{err_n1}, providing an error at $n+1$ 
\begin{equation}
	\varepsilon^{n+1}=\overbrace{\left[\varepsilon^n+A\Delta t^{n+1}\left(\gamma_1\hat{\varepsilon}^n-\gamma_2\hat{\varepsilon}^{n-1}\right)\right]}^{\hat{\varepsilon}^{n+1}}\exp[\underbrace{i\mathbf{k}^n\cdot(\mathbf{x}-\Delta t^{n+1}\mathcal{S}z\mathbf{e}_x)}_{i\mathbf{k}^{n+1}}]\mathrm{.}
  \label{err_n1_rosti}
\end{equation}
As expected, and similarly to what is observed for~\eqref{err_n0}, equation~\eqref{err_n1_rosti} satisfies the condition for $\mathbf{k}^{n+1}$ given by~\eqref{eqn:sec_cond}. This time, however, also the condition~\eqref{eqn:first_cond} is met since the term $\hat{\varepsilon}^{n+1}$ does not contain any spatial-dependent term, i.e., $\mathbf{k}$. The effectiveness of the proposed approach is shown in the next subsection with an analytical benchmark.

\subsection{Validation}
To validate the modified algorithm proposed here, we employ the two-dimensional analytical Kelvin modes derived in the framework of the RDT,
\begin{equation}
\begin{aligned}
  &v^K(\mathbf{x},t) = v_0^K\left(\dfrac{k_0}{k}\right)^2\exp\left[-\dfrac{t\mathcal{S}}{Re_{\lambda}}\left(k_0^2-k_{x,0}k_{y,0}\mathcal{S}t+\dfrac{k_{x,0}^2\mathcal{S}^2t^2}{3}\right)\right]\exp(i\mathbf{k}\cdot\mathbf{x})\mathrm{,} \\
  &u^K(\mathbf{x},t) = -\dfrac{k_y(t)}{k_x}v^K(\mathbf{x},t)\mathrm{,} \\
  &k_y = k_{y,0}-k_x\mathcal{S}t\mathrm{,}
  \label{rdt_2d}
\end{aligned}
\end{equation}
where $Re_{\lambda}$, defined as $\mathcal{S}\lambda_0^2/\nu$, is the Reynolds number based on the perturbation wavelength $\lambda_0=\sqrt{(2\pi/k_x)^2+(2\pi/k_{y,0})^2}$, taken as a reference length. As done by~\cite{kasbaoui2017algorithm}, we consider a two-dimensional square box $l_x=l_y=2\pi$, discretized with $N=128$ points along each direction. The analytical solution~\eqref{rdt_2d} is employed as the initial condition, with initial vertical amplitude $v_0^K=\sqrt{2}/2$, initial wavevector $\mathbf{k}_0=(4,1)$ and $Re_{\lambda}=208$. 
We compare three time-integration methods: (1) the Adams-Bashforth (AB2) in the two variants (classical and modified), (2) the third-order Runge-Kutta (RK3) method proposed in~\cite{tanaka2017effect,yousefi2020modulation} and (3) the Explicit Crank-Nicholson (CN2) method, proposed in~\cite{kasbaoui2017algorithm}. In all cases, the momentum equation is advanced with a constant time step, $\Delta t=0.25\,(\mathcal{S}N)^{-1}$. 
\begin{figure}
    \centering
    \includegraphics[height=4.269 cm,width=12.0 cm]{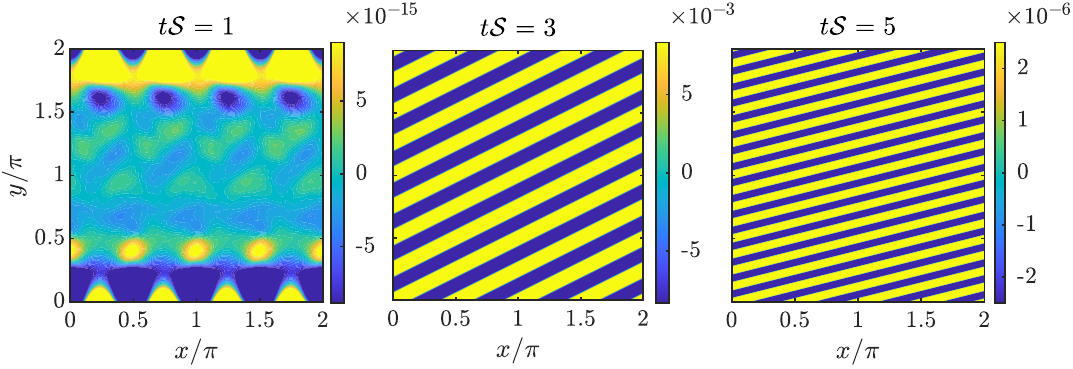}
    \caption{Temporal evolution of the horizontal mode for $t\mathcal{S}=1$, $3$ and $5$ using the modified Adams-Bashforth.}
    \label{fig:ab2_r}
\end{figure} 
\begin{figure}
   \centering
   \includegraphics[height=4.269 cm,width=12.0 cm]{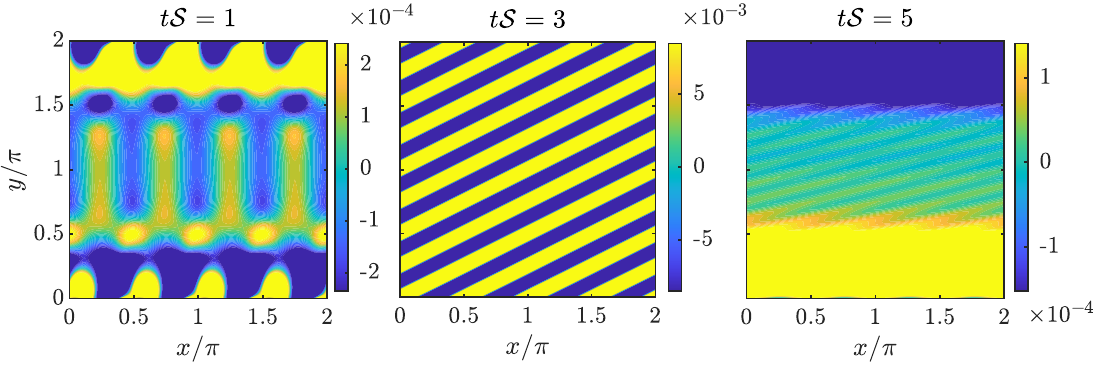}
   \caption{Temporal evolution of the horizontal mode for $t\mathcal{S}=1$, $3$ and $5$ using the standard Adams-Bashforth.}
   \label{fig:ab2_g}
\end{figure}
\begin{figure}
    \centering
    \includegraphics[height=4.269 cm,width=12.0 cm]{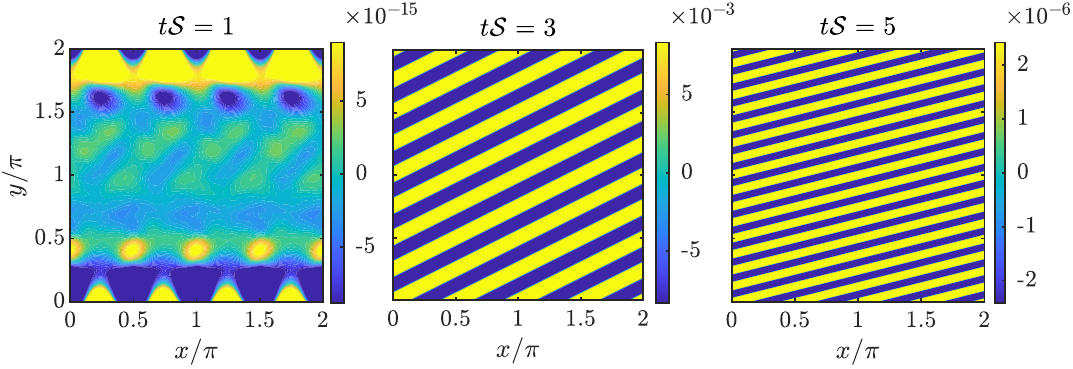}
    \caption{Temporal evolution of the horizontal mode for $t\mathcal{S}=1$, $3$ and $5$ using the explicit Crank-Nicolson method.}
    \label{fig:cn2}
\end{figure}
Figures~\ref{fig:ab2_r}--\ref{fig:cn2} report 
the horizontal mode obtained with the three different time-integration schemes, which according to the RDT should go to zero at $t\mathcal{S}=1$. Using the modified AB2 scheme and CN2, we obtain almost identical results and zero horizontal mode up to machine precision at $t\mathcal{S}=1$. Conversely, using the original approach by Gerz (figure~\ref{fig:ab2_g}), the normal mode does not go to zero for $t\mathcal{S}=1$ and for $t\mathcal{S}=5$ the numerical errors largely affect the solution. \par

\begin{figure}
	\centering
	\includegraphics[width=6.5 cm,height=5.2 cm]{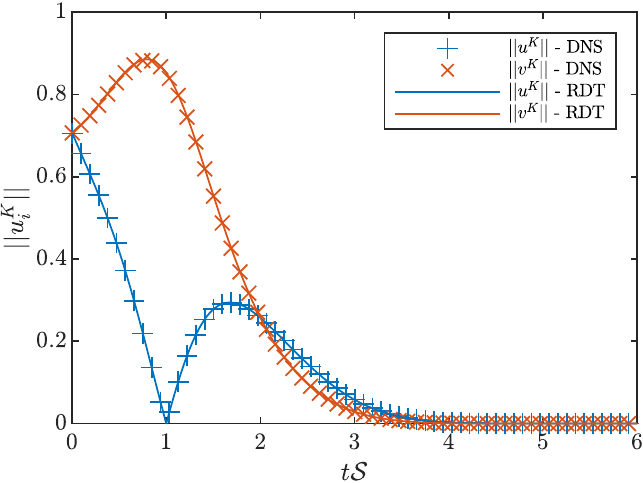}
    \caption{Comparison between the numerical solution obtained with AB2 (proposed method) and the analytical results from RDT.}
	\label{fig:rdt_sol}
\end{figure} 

\begin{figure*}
    \centering
        \includegraphics[width=12 cm,height=5.5 cm]{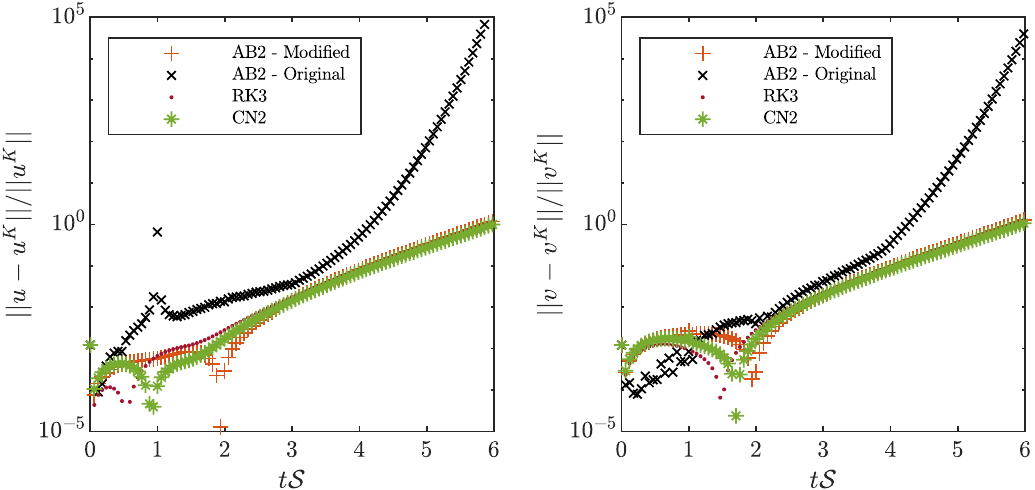}
    ~ 
    \put(-345,140){\small(\textit{a})}
    \put(-165,140){\small(\textit{b})}
    \caption{Temporal evolution of the discretization of error (in log scale) for: (\textit{a}) the horizontal mode (i.e., $u$), (\textit{b}): the vertical mode (i.e., $v$) using the three time integration methods, AB2 - Gerz et al. and AB2 - Modified, RK3  and CN2. }
	\label{fig:error}
\end{figure*}

The excellent agreement between the results obtained with our approach and the analytical solution from RDT are reported in figure~\ref{fig:rdt_sol}. A more detailed analysis considers the evolution in time of the relative error, reported in figure~\ref{fig:error} for the horizontal and vertical Kelvin modes. 
The modified AB2, RK3 and CN2 provide similar errors, especially for $t\mathcal{S}>2$, while for $t\mathcal{S}\leq 2$, the CN2 method appears to be slightly more accurate. On the contrary, the error of the classical AB2 is well above that from the other three methods, with a local peak at $t\mathcal{S}=1$ in the horizontal mode $u^K$, as already observed in the first panel of figure~\ref{fig:ab2_g}. Furthermore, the error evolves at a much faster pace, which is the main reason for the poor performance of the classic AB2 in more demanding simulations.



\bibliographystyle{jfm}
\bibliography{bibfile.bib}

\end{document}